\documentclass[letterpaper,fleqn,usenatbib]{mnras}

\usepackage{newtxtext,newtxmath}

\usepackage[T1]{fontenc}
\usepackage{ae,aecompl}

\usepackage{graphicx}	
\usepackage{amsmath}	
\usepackage{amssymb}	
\usepackage{xcolor}

\title[A Jupiter and Neptune in a triple-star system]{K2-290: a warm Jupiter and a mini-Neptune in a \textcolor{black}{triple-}star system}

\author[M. Hjorth et al.]{M.\,Hjorth$^{1}$\thanks{E-mail: hjorth@phys.au.dk},
A.\,B.\,Justesen$^{1}$,
T.\,Hirano$^{2}$,
S.\,Albrecht$^{1}$,
D.\,Gandolfi$^{3}$,
F.\,Dai$^{4,5}$,
R.\,Alonso$^{6}$,
\newauthor
O.\,Barrag\'an$^{3}$,
M.\,Esposito$^{7}$,
M.\,Kuzuhara$^{8,9}$,
K.\,W.\,F.\,Lam$^{10}$,
J.\,H.\,Livingston$^{11}$,
\newauthor
P.\,Montanes-Rodriguez$^{6}$,
N.\,Narita$^{6,8,9,11}$,
G.\,Nowak$^{6}$,
J.\,Prieto-Arranz$^{6}$, 
S.\,Redfield$^{12}$,
\newauthor
F.\,Rodler$^{13}$,
V.\,Van Eylen$^{5}$,
J.\,N.\,Winn$^{5}$,
G.\,Antoniciello$^{3}$,
J.\,Cabrera$^{14}$,
W.\,D.\,Cochran$^{15}$,
\newauthor
Sz.\,Csizmadia$^{14}$,
J.\,de\,Leon$^{11}$,
H.\,Deeg$^{6,16}$,
Ph.\,Eigm\"uller$^{14}$,
M.\,Endl$^{15}$,
A.\,Erikson$^{14}$,
\newauthor
M.\,Fridlund$^{17,18}$,
S.\,Grziwa$^{19}$,
E.\,Guenther$^{7}$,
A.\,P.\,Hatzes$^{7}$,
P.\,Heeren$^{20}$,
D.\,Hidalgo$^{6,16}$,
\newauthor
J.\,Korth$^{19}$,
R.\,Luque$^{6,16}$,
D.\,Nespral$^{6,16}$,
E.\,Palle$^{6,16}$,
M.\,P\"atzold$^{19}$,
C.\,M.\,Persson$^{17}$,
\newauthor
H.\,Rauer$^{14,21,22}$,
A.\,M.\,S.\,Smith$^{14}$,
T.\,Trifonov$^{23}$\\
$^{1}$Stellar Astrophysics Centre, Department of Physics and Astronomy, Aarhus University, Ny Munkegade 120, DK-8000 Aarhus C, Denmark\\
$^{2}$Department of Earth and Planetary Sciences, Tokyo Institute of Technology, 2-12-1 Ookayama, Meguro-ku, Tokyo 152-8551\\
$^{3}$Dipartimento di Fisica, Universit{\`a} di Torino, via P. Giuria 1, I-10125 Torino, Italy\\
$^{4}$Department of Physics and Kavli Institute for Astrophysics and Space Research, Massachusetts Institute of Technology, Cambridge,\\ MA 02139, USA\\
$^{5}$Department of Astrophysical Sciences, Princeton University, 4 Ivy Lane, Princeton, NJ 08544, USA\\
$^{6}$Instituto de Astrof{\'i}sica de Canarias, C/ V{\'i}a L{\'a}ctea s/n, E-38205 La Laguna, Spain\\
$^{7}$Th{\"u}ringer Landessternwarte Tautenburg, Sternwarte 5, D-07778 Tautenberg, Germany\\
$^{8}$Astrobiology Center, NINS, 2-21-1 Osawa, Mitaka, Tokyo 181-8588, Japan\\
$^{9}$National Astronomical Observatory of Japan, NINS, 2-21-1 Osawa, Mitaka, Tokyo 181-8588, Japan\\
$^{10}$Zentrum f\"ur Astronomie und Astrophysik, Technische Universit\"at Berlin, Hardenbergstr. 36, 10623 Berlin, Germany\\
$^{11}$Department of Astronomy, Graduate School of Science, The University of Tokyo, Hongo 7-3-1, Bunkyo-ku, Tokyo, 113-0033, Japan\\
$^{12}$Astronomy Department and Van Vleck Observatory, Wesleyan University, Middletown, CT 06459, USA\\
$^{13}$European Southern Observatory, Alonso de C{\'o}rdova 3107, Vitacura, Casilla, 19001 Santiago de Chile, Chile\\
$^{14}$Institute of Planetary Research, German Aerospace Center, Rutherfordstrasse 2, 12489 Berlin, Germany\\
$^{15}$Department of Astronomy and McDonald Observatory, University of Texas at Austin, 2515 Speedway,~Stop~C1400,~Austin,~TX~78712,~USA\\
$^{16}$Departamento de Astrof\'isica, Universidad de La Laguna, E-38206, Tenerife, Spain\\
$^{17}$Department of Space, Earth and Environment, Chalmers University of Technology, Onsala Space Observatory, 439 92 Onsala, Sweden\\
$^{18}$Leiden Observatory, Leiden University, 2333CA Leiden, The Netherlands\\
$^{19}$Rheinisches Institut f\"ur Umweltforschung an der Universit\"at zu K\"oln, Aachener Strasse 209, 50931 K\"oln, Germany\\
$^{20}$ZAH-Landessternwarte Heidelberg, K{\"o}nigstuhl 12, D-69117 Heidelberg, Germany\\
$^{21}$Institute of Geological Sciences, FU Berlin, Malteserstr. 74-100, D-12249 Berlin, Germany\\
$^{22}$Center for Astronomy and Astrophysics, TU Berlin, Hardenbergstr. 36, 10623 Berlin, Germany\\
$^{23}$Max-Planck-Institut f{\"u}r Astronomie, K{\"o}nigstuhl 17, D-69117 Heidelberg, German\\
\vspace{-1.2cm}
}

\date{Accepted 2018 December 20. Received 2018 December 20; in original form 2018 October 02}

\pubyear{2018}

\begin{document}
\label{firstpage}
\pagerange{\pageref{firstpage}--\pageref{lastpage}}
\maketitle

\begin{abstract}
We report the discovery of two transiting planets orbiting K2-290 (EPIC\,249624646), a bright (V=11.11) late F-type star \textcolor{black}{residing in a triple-star system}. 
It was observed during Campaign 15 of the {\it K2} mission, and in order to confirm and characterise the system, follow-up spectroscopy and AO imaging were carried out using the FIES, HARPS, HARPS-N, and IRCS instruments. From AO imaging and {\it Gaia} data we identify two M-dwarf companions at a separation of $113 \pm 2$~AU and $2467_{-155}^{+177}$~AU. From radial velocities, {\it K2} photometry, and stellar characterisation of the host star, we find the inner planet to be a mini-Neptune with a radius of $3.06 \pm 0.16$~R$_{\earth}$ and an orbital period of $P = 9.2$~days. The radius of the mini-Neptune suggests that the planet is located above the radius valley, and with an incident flux of $F\sim 400$~F$_{\earth}$, it lies safely outside the super-Earth desert. The outer warm Jupiter has a mass of \textcolor{black}{$0.774\pm 0.047 M_{\rm J}$} and a radius of \textcolor{black}{$1.006\pm 0.050R_{\rm J}$}, and orbits the host star every 48.4~days on \textcolor{black}{an orbit with an eccentricity $e<0.241$}. Its mild eccentricity and mini-Neptune sibling suggest that the warm Jupiter originates from {\it in situ} formation or disk migration.

\end{abstract}




\section{Introduction}
With the success of the {\it Kepler} mission \citep{Kepler}, exoplanetary science entered a new era. With the breakdown of its second reaction wheel in 2013, the spacecraft continued operating through the {\it K2} mission \citep{Howell2014}. Because of its monitoring of fields at the ecliptic in timeslots of $\sim80$~days, the $K2$ mission has been able to target an area of the sky, which will have limited coverage in the {\it TESS} mission \citep{TESS}. Combined, {\it Kepler} and {\it K2} have to date discovered more than 2500 confirmed planets\footnote{\url{https://nasa.gov/mission_pages/kepler}} -- an essential achievement for our understanding of these new worlds.

Of the large number of exoplanets discovered, some are very different from the Solar System planets. This is e.g. the case for super-Earths and mini-Neptunes, which have sizes between Earth and Neptune, and for hot and warm Jupiters, which are Jupiter-sized planets with orbital periods of $<10$~days and between $10$ and $\sim200$~days, respectively. Our understanding of the formation of these planets is still limited. In the case of hot Jupiters it appears as if they formed at significantly larger orbits than where we find them now\footnote{see however \citet{Batygin2016} for specific scenarios of {\it{in situ}} formation.}, but their migration mechanism(s) is yet to be determined (see \citet{Dawson2018} for a review). Planetary migration via angular momentum exchange with the protoplanetary disk \citep[e.g.][]{Lin1996} would lead to low eccentricity orbits roughly aligned with the disk midplane. Whereas high-eccentricity migration \citep[e.g.][]{RasioFord1996} would lead to large eccentricities ($\gtrsim0.2$) and orbits outside the disk midplane. Interpretation of these orbital parameters in the framework of planet formation and migration is however complicated by tidal damping of orbital eccentricities \citep[e.g.][]{Bonomo2017} and by tidal alignment of orbital and stellar spins \citep{Winn2010,Albrecht2012}, the latter being under debate (see \citet{ZanazziAndLai2018} and references therein).

Some warm Jupiters might be progenitors of hot Jupiters, but their orbits will be altered less by tidal damping due to the larger separations from the host stars \citep{Petrovich2016}.  
In addition, studying the eccentricity \citep{Dong2014} and companionship \citep{Huang2016} of warm Jupiter systems, it has been proposed that warm Jupiters originate from two different formation paths: high-eccentricity migration (i.e. as hot Jupiter progenitors) and {\it{in situ}} formation. If they originate from high-eccentricity migration these are predicted to have undergone secular eccentricity oscillations by the hand of an outer close-by high-mass companion and have high eccentricities \citep[$>0.4$;][]{Dong2014, Petrovich2016} and no low-mass inner companions \citep{2015ApJ...808...14M}, while if they form {\it{in situ}} they should have low eccentricities \citep[$<0.2$;][]{Petrovich2016} and inner low-mass siblings with low mutual inclinations \citep{Huang2016}. Determinating companionship and orbital eccentricities should therefore shed light on the origin of both hot and warm Jupiters \citep[see][and references therein]{Dawson2018}.

In the case of warm and hot Jupiters forming through dynamical pertubations of their orbits, the formation might be somewhat more efficient within a triple star system than in binary star systems \citep[see][and references therein]{Hamers2017}. However, only a couple dozens of planetary systems have been confirmed to be in triple star systems\footnote{\href{univie.ac.at/adg/schwarz/bincat_multi_star.html}{Catalogue} introduced in \citet{Schwarz2016}}. We are only aware of two of these having multiple planets: GJ 667C \citep{Anglada2012, Feroz2014} and Kepler-444A \citep{Campante2015}, both of which contain no giant planets.

Here we present the discovery, confirmation and characterisation of the multitransiting planet system K2-290 (EPIC\,249624646) detected by the {\it K2} mission. K2-290b is a mini-Neptune on a $\sim9.2$~day orbit, while K2-290c is a warm Jupiter with an orbital period of $\sim48.4$~days. They both orbit the bright late F-type sub-giant K2-290 ($V=11.11$), which in turn have two stellar companions, probably as a member of a triple-star system. We used a combination of {\it Kepler} photometry, high-resolution spectrocopy from FIES, HARPS and HARPS-N and AO imaging from IRCS to detect and characterise the planets and their orbits. This was done as part of the KESPRINT collaboration\footnote{\url{https://iac.es/proyecto/kesprint/}}, which aims to confirm and characterise {\it K2} \textcolor{black}{and {\it TESS}} systems \citep[see e.g.][]{2018MNRAS.478.4866V, 2018AJ....156...78L, 2018MNRAS.481..596J}.

The paper is structured in the following way: In section\,\ref{sec: obs} the observational data consisting of photometry, spectroscopy and AO imaging are presented. The analysis of the host star and its two companions is presented in section\,\ref{sec:star}, while section\,\ref{sec:planet} deals with the planetary confirmation and characterisation of K2-290b and K2-290c. In section\,\ref{sec:discussion} our findings are discussed and put into context.

\section{Observations}
\label{sec: obs}

To detect, characterise and analyse the planets and stars in the system, we use several different types of observations. This includes photometry, high-resolution spectroscopy, and AO imaging. An overview of the data sources and data characteristics can be found in Table\,\ref{table:obs_overview}. A detailed description of the observations are given in this section.

\begin{table}
\begin{center}
\centering
\caption{Observation log of K2-290 containing the different types of observation, instrument, instrument resolution, no. of observations made and observing dates. {\bf{Notes}}:$^{1}$The original no. of observations.}
\label{table:obs_overview}
\resizebox{\columnwidth}{!}{%
\begin{tabular}{lcccc}
\hline\hline
\noalign{\smallskip}
Type & Inst. & Spec. res. & No. of obs. & Obs. date\\
\noalign{\smallskip}
\hline
Phot. & {\it{Kepler}} & -- & 3909$^{1}$ & 2017 8/23 -- 11/20\\
\noalign{\smallskip}
\hline
{} & HARPS & 115000 & 16 & 2018 2/23 -- 5/12\\
Spec. & HARPS-N & 115000 & 6 & 2018 2/20 --7/14\\
{} & FIES & 47000 & 11 & 2018 5/12 -- 7/13\\
\noalign{\smallskip}
\hline
\textcolor{black}{Imaging} & IRCS & -- & 2 & 2018 3/29 \textcolor{black}{\&} 6/14
\end{tabular}%
}
\end{center}
\end{table}

\subsection{K2 photometry}
The star K2-290 was observed by the Kepler space telescope in Campaign 15\footnote{\textcolor{black}{Guest observer programmes GO15009\_LC, GO15021\_LC, GO028\_LC and GO083\_LC.}} of the {\it K2} mission \citep{Howell2014}. A total of 3909 long-cadence observations (29.4 min integration time) were made of this target between August 23 and November 20 2017. For a detailed analysis, we downloaded the pre-processed lightcurve from MAST\footnote{\url{https://archive.stsci.edu/prepds/k2sff}}, which is reduced from the raw data following the procedure described in \cite{Vanderburg2014}.
The search method for transiting exoplanet candidates in the {\it{K2}} data is described in \citet{Dai2017}, which follows a similar approach as \cite{Vanderburg2014}.

Two transit signatures were detected in the lightcurve of K2-290 with periods of $\sim9.2$~days and $\sim48.4$~days and depths of $\sim0.03 \%$ and $\sim0.5 \%$, respectively (see Fig.\,\ref{fig:lc}). This is consistent with a mini-Neptune or super-Earth and a warm Jupiter orbiting a slightly evolved F8 star.

The out-of-transit signal is fairly quiet: we find no evidence of recurring stellar spots and in general no signs of any additional periodic signals in the lightcurve.

\subsection{Spectroscopy}
\label{sec:spectroscopy}
\begin{figure*}
\centering
\includegraphics[width=\columnwidth]{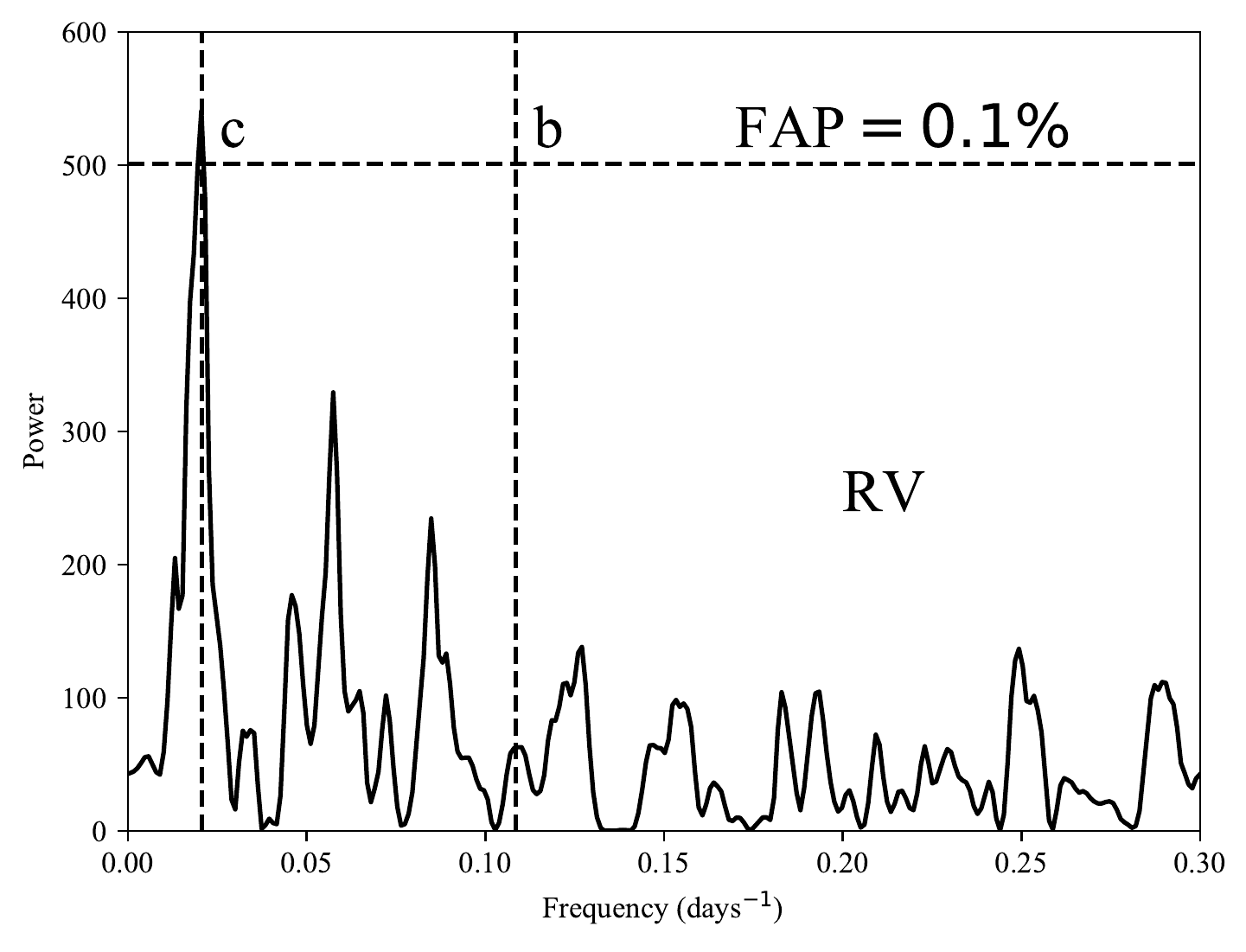}
\includegraphics[width=\columnwidth]{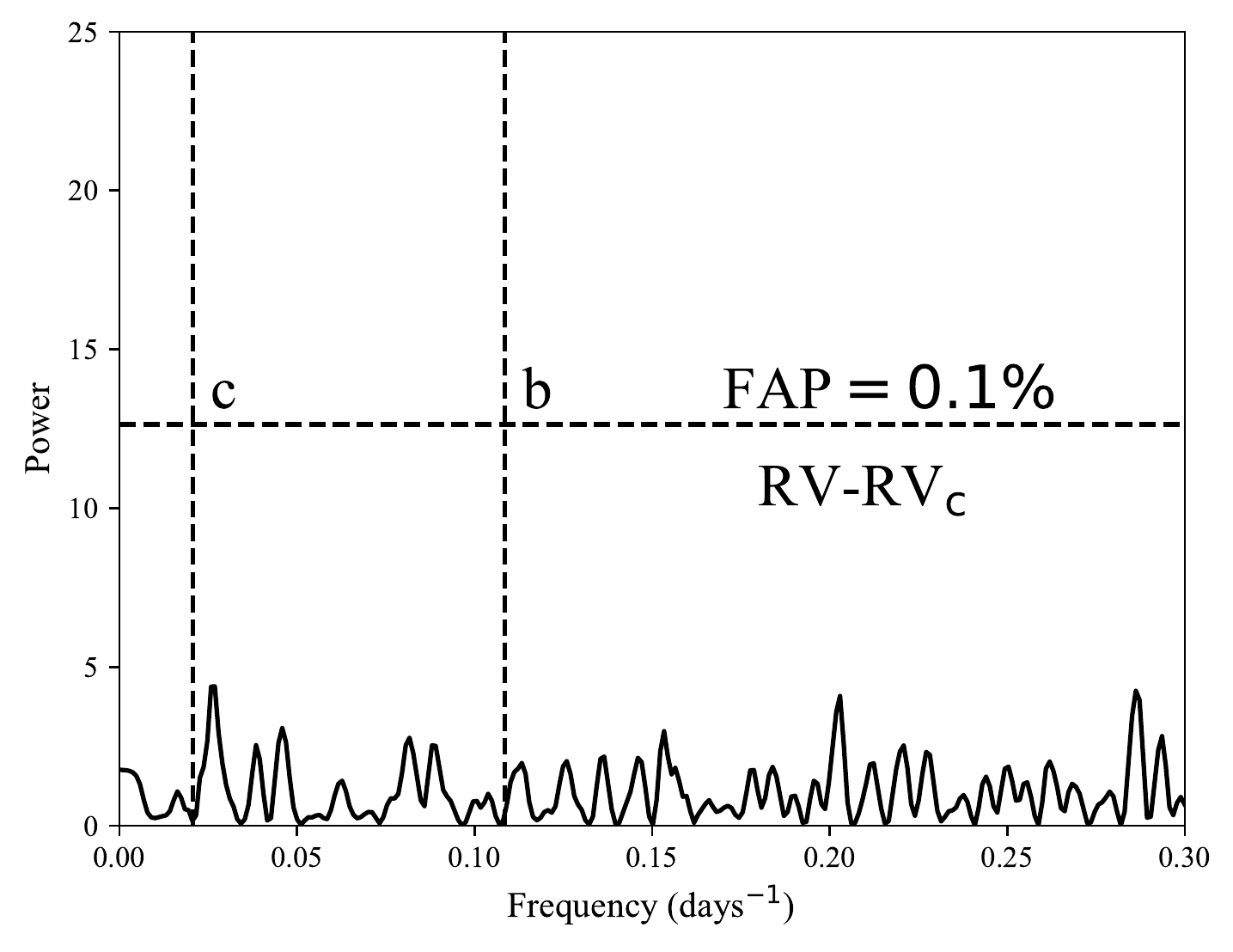}
\caption{The GLS periodograms of the RVs using offsets subtracted data only (left) and additionally the best-fitting keplerian model for planet c subtracted (right). The dashed vertical lines mark the frequencies at which we expect to find the signals for planet b and c, given the orbital periods from the photometric data. The dashed horizontal lines indicate the respective $0.1\%$ false alarm probabilities.}
\label{fig:LS}
\end{figure*}
Spectroscopic observations of K2-290 were carried out between 2018/02/20 and 2018/08/28 using the FIES, HARPS, and HARPS-N spectrographs.

The FIES (Fiber-fed Echelle Spectrograph; \citealp{FIES2014}) spectra were gathered between 2018/05/12 and 2018/07/13 at the 2.56~m Nordic Optical Telescope (NOT) of Roque de los Muchachos Observatory, La Palma, Spain. We obtained 11 med-resolution spectra ($R\sim47000$) as part of the Nordic and OPTICON programmes 57-015 and 2018A/044, using the observing strategy described in \citet{Gandolfi2013}. The spectra were reduced using standard IRAF and IDL\footnote{\url{https://idlastro.gsfc.nasa.gov/}} routines, and radial velocities (RVs) were extracted through fitting Gaussians to multi-order cross-correlation functions (CCFs) using the stellar spectrum with the highest S/N as template.

Between February 23 and August 28 2018, we also obtained 16 high-resolution ($R \sim 115000$) spectra with the High Accuracy Radial velocity Planet Searcher spectrograph \citep[HARPS;][]{Mayor2003} mounted at the ESO 3.6~m telescope of La Silla observatory. The spectra were gathered in connection with the ESO programmes 0100.C-0808 and 0101.C-0829. The data were reduced using the offline HARPS pipeline. The RVs were extracted through cross-correlations of the processed spectra with a G2 numerical mask \citep{Pepe2002}.

We further used the HARPS-N spectrograph \citep{HARPSN} installed at the 3.6~m Telescopio Nazionale Galileo (TNG) of the Roque de los Muchachos Observatory, La Palma, Spain. Here we collected 6 high-resolution ($R \sim 115000$) spectra between 2018/02/20 and 2018/07/14 as part of the Spanish and TAC programmes CAT17B{\_}99, CAT18A{\_}130, and A37TAC{\_}37. The data were reduced and RVs extracted using the same procedure as done for the HARPS data.

In total, 33 spectra were obtained and reduced. In Table\,\ref{table:RVs} we list the barycentric time of mid-exposure, the RVs, the RV uncertainties ($\sigma_{RV}$), the bisector span (BIS) and the full-width at half maximum (FWHM) of the CCFs, the exposure times, the signal-to-noise ratios (S/N) per pixel at 5500 \AA, and the instrument used for a specific observation.

We performed a frequency analysis of the RV measurements to test whether the two transiting planet candidates are detectable in the spectroscopic data. This was done by computing the generalized Lomb-Scargle (GLS) periodogram \citep{LS} of the combined FIES, HARPS, and HARPS-N measurements. The RV data were first corrected for the instrument offsets using the values derived from the global analysis described in Sec.\,\ref{subsub:global}. The GLS periodogram (Fig.~\ref{fig:LS}, left panel) shows a significant peak at the orbital frequency of planet c (false alarm probability FAP\,$<$\,0.1$\%$, calculated using the bootstrap method from \citealp{1997AA...320..831K}), indicating that we would have been able detect planet c even in the absence of the K2 photometry. However, we do not see a significant peak at the frequency of planet b. Even subtracting the best-fitting Keplerian model for planet c (from the analysis described in Sec.\,\ref{subsub:global}), we see no signs of its small sibling (Fig.~\ref{fig:LS}, right panel).

\subsection{AO Imaging}

\begin{figure*}
\centering
\includegraphics[width=\columnwidth]{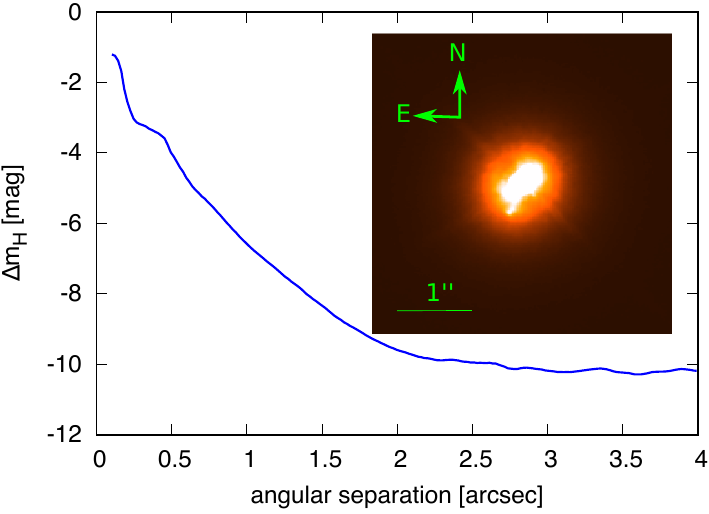}
\includegraphics[width=\columnwidth]{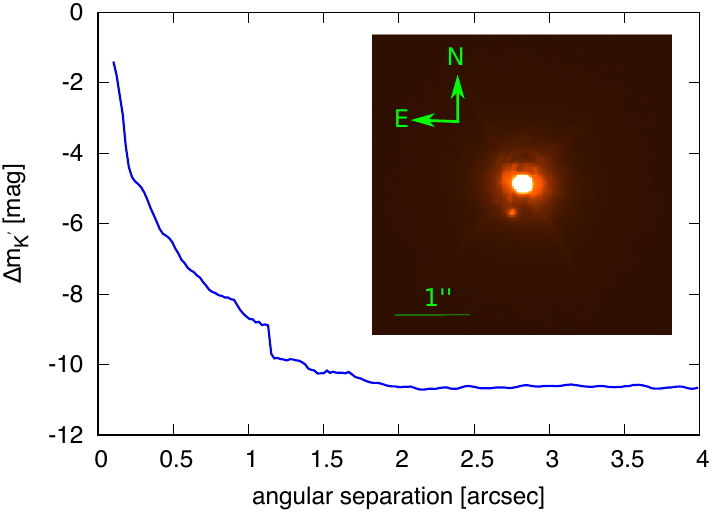}
\caption{The 5-$\sigma$ contrast curves and 4$''\times$4$''$ Field-of-view AO images (inset) in the $H$ band (left) and $K'$ band (right) for observations done with the IRCS at the Subaru Telescope. With K2-290 in the center, the images reveal a faint neighbouring star about 0.4$''$ away.}
\label{fig:AO}
\end{figure*}
We conducted adaptive optics (AO) imaging using IRCS (Infrared Camera and Spectrograph; \citealp{IRCS, IRCSAO}) on the 8.2~m Subaru Telescope at the Mauna Kea Observatory, Hawaii, US as part of the programme S18A-089. With these observations we aimed at ruling out a false positive transit signal caused by an eclipsing binary as well as to search for potential stellar companions of K2-290. We obtained $H$ band observations on March 29 2018 and $K'$ band observations on June 14 2018. For both observing bands we executed two sequences: one for saturated frames and the other for unsaturated, with a five-point dithering. Since the target image becomes saturated with the shortest integration ($<1$\,s), we used a neutral-density (ND) filter (transmittance $\sim 1\%$) for unsaturated frames. The total integration times for the saturated frames were 75 s and 37.5 s, for the $H$ and $K^\prime$ bands, respectively. We used the target star itself as a natural guide star for AO. The images in both bands were reduced following the procedure described in \citet{Hirano2016}. We describe our procedures for contrast analysis and aperture photometry in Sec.\,\ref{subsec:companions}. The contrast curves and reduced AO images in both the $H$ and $K'$ bands are inset in Fig. \ref{fig:AO}. We note that the central part of the H band image is saturated \textcolor{black}{and that it clearly displays a deformation. While we have not been able to pinpoint the exact cause, we assume here that it is related to the instrument or sky condition. However, the photometry uncertainty caused by this deformed PSF can be mitigated since we performed a relative photometry between the companion's Point Spread Functions (PSFs) observed in the saturated images and the parent star's PSF observed in the unsaturated images, which were obtained soon before the saturated frames.}

The resulting AO images hints at the presence of a possible companion only $\sim0.4''$ away and reveals another possible companion star at a distance of $\sim10''$ (not displayed in the image, see Sec.\,\ref{subsec:companions}).
From now on, the potential inner companion will be referred to as star B, and the potential outer companion as star C.

\section{Stellar characterisation}
\label{sec:star}
\subsection{Host star properties}
\begin{table}
\begin{center}
\centering
\caption{Identifiers, coordinates, kinematics and magnitudes of the host star K2-290. EPIC is the Ecliptic Plane Input Catalogue (\url{https://archive.stsci.edu/k2/epic/search.php}), while {\it Gaia} refer to parameters extracted from {\it Gaia} DR2 (\citealp{Gaia2018}, \url{https://gea.esac.esa.int/archive/}). Besides the {\it{Kepler}} magnitude, the magnitudes from EPIC are collected from \citet{Tycho2} and \citet{2MASS}. \textcolor{black}{\textbf{Notes:} *}As discussed in Sec.\,\ref{sec:false_pos}, the \textcolor{black}{literature} magnitudes reflect the {\it{combined}} magnitudes of the host star and star B. \textcolor{black}{$^{\dagger}$Obtained from estimated {\it{Sloan}} $r$ and $g$ magitudes for star B, converted to {\it{Kepler}}-magnitude using \citet{Brown2011} (see Sec. \ref{sec:false_pos}). $^{\ddagger}$Assuming the $K'$-band of IRCS is equal to the $K$-band of {\it{2MASS}}.}}
\label{table:host_star_parameters}
\resizebox{\columnwidth}{!}{%
\begin{tabular}{lcc}
\hline\hline
\noalign{\smallskip}
Parameter                         & Value & Source \\
\noalign{\smallskip}
\hline
K2                                        & 290 &  \\
EPIC                                        & 249624646 & EPIC \\
TYC                                        & 6193-663-1 & EPIC \\
{\it Gaia} DR2                                        & \textcolor{black}{6253844468882760832} & {\it Gaia} \\
\noalign{\medskip}
$\alpha$ (J2000.0)                                        & 15$^{\text{h}}$ 39$^{\text{m}}$ 25.865$^{\text{s}}$                            & EPIC                            \\
$\delta$ (J2000.0)                                        & -20$^{\circ}$ 11$^{\text{m}}$ 55.74$^{\text{s}}$                            & EPIC                            \\
parallax (mas)                                        & 3.636$\pm 0.050$                            & {\it Gaia}                            \\
distance (pc)                            & 275.0$\pm 3.8$                            & {\it{Gaia}}                            \\
\textcolor{black}{systemic RV (km s$^{-1}$)}                            & \textcolor{black}{19.70$\pm 0.37$}                            & \textcolor{black}{\it{Gaia}}                            \\
$\mu_{\alpha}$ (mas yr$^{-1}$)                                        & 27.225$\pm 0.099$                            & {\it Gaia}                            \\
$\mu_{\delta}$ (mas yr$^{-1}$)                                        & -16.893$\pm 0.066$                            & {\it Gaia}                            \\
\noalign{\medskip}
{\it \textcolor{black}{Combined mag.*}}                         & \multicolumn{1}{l}{} & \multicolumn{1}{l}{} \\
$G$                                        & 10.8204$\pm 0.0004$                            & {\it Gaia}                            \\
{\it{Kepler}}                                   & 10.784                           & EPIC                            \\
$B$                                        & 11.68$\pm 0.11$                            & EPIC                            \\
$V$                                        & 11.11$\pm 0.11$                           & EPIC                            \\
$J$                                        & 9.771$\pm 0.022$                           & EPIC                            \\
$H$                                        & 9.477$\pm 0.022$                            & EPIC                            \\
$K$                                        & 9.420$\pm 0.019$                           & EPIC                            \\
$g$                                        & 11.179$\pm 0.030$                           & EPIC                            \\
$r$                                        & 10.784$\pm 0.030$                           & EPIC                            \\
$i$                                        & 10.614$\pm 0.020$                           & EPIC                            \\
\noalign{\medskip}
{\it \textcolor{black}{Derived host star mag.}}                         & \multicolumn{1}{l}{} & \multicolumn{1}{l}{} \\
{\it{\textcolor{black}{Kepler$^{\dagger}$}}}                                   & \textcolor{black}{10.785}                           & \textcolor{black}{This work}                            \\
\textcolor{black}{$H$}                                        & \textcolor{black}{9.494$\pm 0.022$}                            & \textcolor{black}{This work}                            \\
\textcolor{black}{$K^{\ddagger}$}                                        & \textcolor{black}{9.441$\pm 0.019$}                           & \textcolor{black}{This work}                            \\
\noalign{\medskip}
{\it Derived parameters}                         & \multicolumn{1}{l}{} & \multicolumn{1}{l}{} \\
$M_{\star}$ ($M_{\sun}$)                            & $1.194^{+0.067}_{-0.077}$                            & This work                            \\
$R_{\star}$ ($R_{\sun}$)                          & $1.511^{+0.075}_{-0.072}$                           & This work                            \\
$\rho_{\star}$ (g cm$^{-3}$)                     & $0.485^{+0.074}_{-0.064}$                            & This work                            \\
$T_{\text{eff}, \star}$ (K)                       & $6302 \pm 120$                            & This work                            \\
$\log {\text{g}}_{\star}$ (cgs)                       & $4.23 \pm 0.10$                            & This work                            \\
$v\sin i_{\star}$ (km s$^{-1}$)                       & $6.5 \pm 1.0$                            & This work                            \\
$[$Fe/H$]$ (dex)                       & $-0.06 \pm 0.10$                            & This work                            \\
age (Gyr)                       & $4.0^{+1.6}_{-0.8}$                            & This work                            \\

\end{tabular}%
}
\end{center}
\end{table}
In the first step of the data analysis, we aimed to determine the absolute stellar parameters of K2-290. To this end, we created a high signal-to-noise (S/N) spectrum by co-adding the individual HARPS spectra having S/N ratios of ~60 per spectral pixel at 5500 \AA{} (see Table~\ref{table:RVs}). \textcolor{black}{This resulted in a co-added spectrum with a total S/N of $\sim150$.}
We then used the iSpec framework \citep{2014A&A...569A.111B} to fit synthetic stellar spectra computed using the SYNTHE \citep{1993sssp.book.....K} with MARCS model atmospheres \citep{2008A&A...486..951G} to the high S/N spectrum. \textcolor{black}{We assumed a Gaussian spectral PSF with a FWHM corresponding to $R = 115000$ over the spectral bandpass of the HARPS spectrograph.} We fitted the effective stellar temperature ($T_{\text{eff}}$), surface gravity ($\log g$), metallicity ([Fe/H]) and projected stellar rotation speed ($v \sin i_\star$), while fixing the micro- and macroturbulence parameters ($v_{\rm mic}$ and $v_{\rm mac}$). \textcolor{black}{We fixed $v_{\rm mic}=1.3$~km~s$^{-1}$ and $v_{\rm mac}=5.0$~km~s$^{-1}$ using the empirical relations calibrated for the \textit{Gaia}-ESO Survey as implemented in iSpec \citep{2014A&A...569A.111B}. We have tested that fixing $v_{\rm mic}$ and $v_{\rm mac}$ do not significantly affect the derived spectroscopic parameters compared to keeping them free. Macroturbulence $v_{\rm mac}$ and rotational broadening $v \sin i_\star$ are degenerate at the resolution and S/N of our spectrum. The choice of $v_{\rm mac}$ therefore affects $v \sin i_\star$, but not other quantities. $v_{\rm mic}$ is similarly difficult to determine accurately from the spectrum, but only affects other quantities weakly. We have reanalysed the spectrum while keeping $v_{\rm mic}$ and $v_{\rm mac}$ free, and find parameters that agree within their uncertainties. After carrying out the fit} we combined the information extracted from our spectroscopic analysis ($T_{\text{eff}}$, $\log g$, and [Fe/H]) with the \textit{Gaia} DR2 parallax \citep{Gaia2018} and apparent magnitude in the $H$-band (corrected for the contamination of the close companion, see Sec. \ref{sec:false_pos}). For the parallax error, 0.1~mas is added in quadrature to account for systematic uncertainties \citep{Luri2018}. We estimate \textcolor{black}{an} interstellar reddening using the dust map by \citet{Green2018}. Reddening is transformed into extinction in the $H$-band using the relations by \citet{CasagrandeVanderberg2014, CasagrandeVanderberg2018}. Using the recently updated isochrones from the BaSTI database \citep{Hidalgo2018} and the BAyesian STellar Algorithm {\tt BASTA} \citep{Victor2015} we obtain a stellar mass of $1.19^{+0.07}_{-0.08}$ $M_{\sun}$, a radius of $1.51^{+0.08}_{-0.07}$ $R_{\sun}$, and an age of $4.0^{+1.6}_{-0.8}$ Gyr. See Table\,\ref{table:host_star_parameters} for a complete listing of the parameters. \textcolor{black}{As a consistency check we use the reddening- and contamination-corrected $V$ magnitude, the \textit{Gaia} DR2 parallax and the spectroscopic $T_{\text{eff}}$ to determine the stellar radius using the \citet{2010AJ....140.1158T} bolometric correction. We derive a radius $R_* = 1.42 \pm 0.1 R_{\sun}$, in agreement within $1 \sigma$ of the radius derived using \texttt{BASTA}.}

\subsection{Stellar companions}
\label{subsec:companions}
\begin{figure}
\centering
\includegraphics[width=\columnwidth]{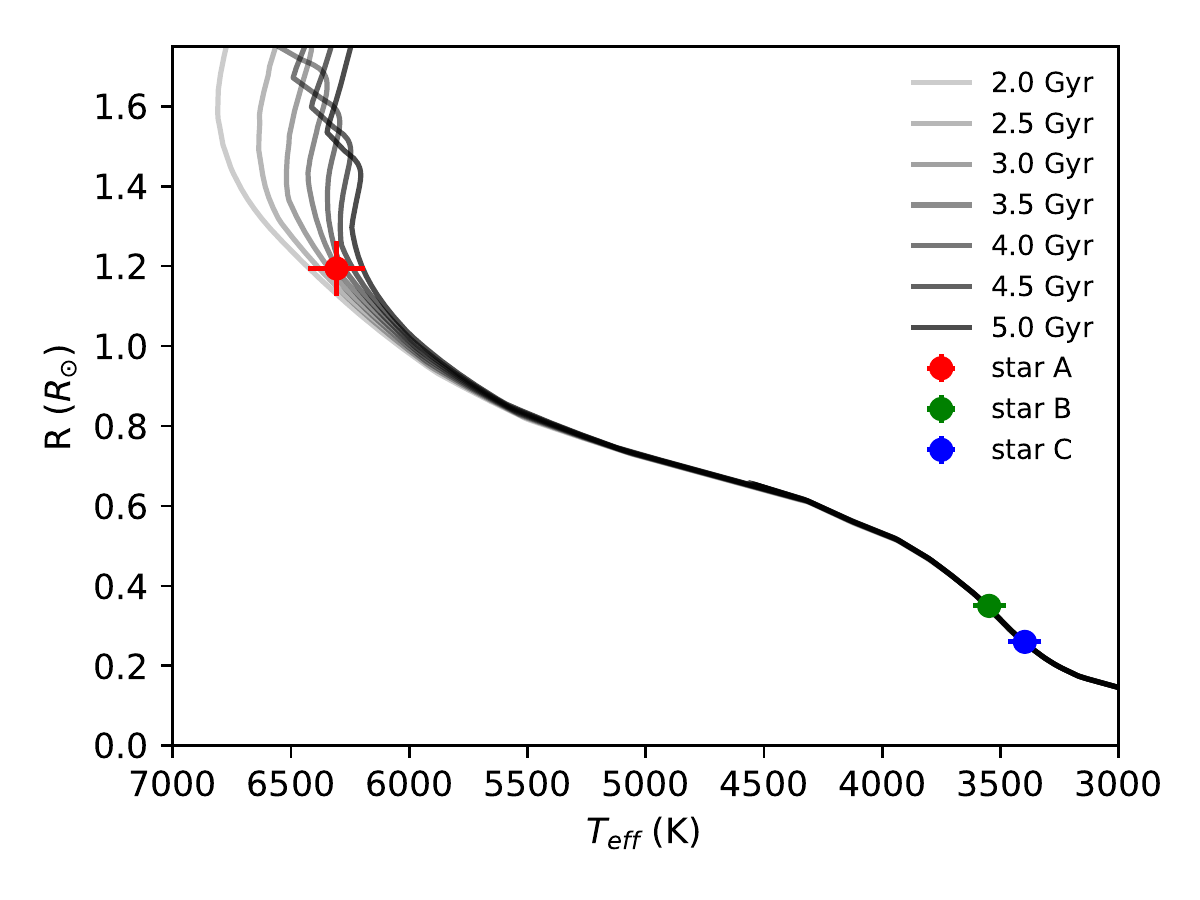}
\caption{The H-R diagram for K2-290 (star A) and its two stellar companions (star B and star C) together with BaSTI isochrones ranging from 2.0 to 5.0 Gyr and with $[$Fe/H$] = -0.1$.}
\label{fig:HR}
\end{figure}

\begin{table}
\begin{center}
\centering
\caption{Available identifiers, coordinates, kinematics and magnitudes of the two stellar companions to K2-290, together with derived parameters from analysis of the AO images. {\it Gaia} refer to parameters extracted from {\it Gaia} DR2 (\citealp{Gaia2018}, \url{https://gea.esac.esa.int/archive/}), while {\it 2MASS} magnitudes are from \citet{2MASS}. 
\textcolor{black}{\textbf{Notes:} *Obtained from estimated {\it{Sloan}} $r$ and $g$ magitudes for star B, converted to {\it{Kepler}}-magnitude using \citet{Brown2011} (see Sec. \ref{sec:false_pos}). $^{\dagger}$Assuming the $K'$-band of IRCS is equal to the $K$-band of {\it{2MASS}}.}
}
\label{table:companion_parameters}
\resizebox{\columnwidth}{!}{%
\begin{tabular}{lcc}
\hline\hline
\noalign{\smallskip}
Parameter                         & Value & Source \\
\noalign{\smallskip}
\hline
{\bf Star B (Close-by component)}                         & \multicolumn{1}{l}{} & \multicolumn{1}{l}{} \\
\noalign{\medskip}
{\it AO imaging $H$-band}                         & \multicolumn{1}{l}{} & \multicolumn{1}{l}{} \\
$\Delta H$                                        & 4.474$\pm 0.092$                            & This work                            \\
ang. sep. (arcsec)                                        & 0.389$\pm 0.008$                            & This work                            \\
pos. angle (degree)                                        & 160.1$\pm 1.4$                            & This work                            \\
\noalign{\medskip}
{\it AO imaging $K'$-band}                         & \multicolumn{1}{l}{} & \multicolumn{1}{l}{} \\
$\Delta K'$                                       & \textcolor{black}{4.270$\pm 0.036$}                            & This work                            \\
ang. sep. (arcsec)                                        & 0.411$\pm 0.015$                            & This work                            \\
pos. angle (degree)                                        & 159.2$\pm 2.8$                            & This work                            \\
\noalign{\medskip}
{\it \textcolor{black}{Derived mag.}}                         & \multicolumn{1}{l}{} & \multicolumn{1}{l}{} \\
{\it{\textcolor{black}{Kepler*}}}                                   & \textcolor{black}{17.981}                           & \textcolor{black}{This work}                            \\
\textcolor{black}{$H$}                                        & \textcolor{black}{13.968$\pm 0.093$}                            & \textcolor{black}{This work}                            \\
\textcolor{black}{$K^{\dagger}$}                                        & \textcolor{black}{13.711$\pm 0.040$}                           & \textcolor{black}{This work}                            \\
\noalign{\medskip}
{\it Derived parameters}                         & \multicolumn{1}{l}{} & \multicolumn{1}{l}{} \\
$M_{\text{B}}$ ($M_{\sun}$)                            & $0.368 \pm 0.021$                            & This work                            \\
$R_{\text{B}}$ ($R_{\sun}$)                          & $0.354 \pm 0.017$                            & This work                            \\
$T_{\text{eff},\text{B}}$ (K)                       & $3548 \pm 70$                            & This work                            \\

\hline
{\bf Star C (Far away component)}                         & \multicolumn{1}{l}{} & \multicolumn{1}{l}{} \\
\noalign{\medskip}
{\it Gaia} DR2                                        & 6253844464585162880                            & {\it Gaia}                            \\
\noalign{\medskip}
$\alpha$ (J2000.0)                                        & 15$^{\text{h}}$ 39$^{\text{m}}$ 28.390$^{\text{s}}$                            & {\it Gaia}                            \\
$\delta$ (J2000.0)                                        & -20$^{\circ}$ 12$^{\text{m}}$ 7.282$^{\text{s}}$                            & {\it Gaia}                            \\
parallax (mas)                                        & 4.053$\pm 0.271$                            & {\it Gaia}                            \\
distance (pc)                            & 247$^{+18}_{-16}$                            & {\it{Gaia}}                            \\
$\mu_{\alpha}$ (mas yr$^{-1}$)                                        & 27.224$\pm 0.099$                            & {\it Gaia}                            \\
$\mu_{\delta}$ (mas yr$^{-1}$)                                        & -16.484$\pm 0.370$                            & {\it Gaia}                            \\
\noalign{\medskip}
$G$                                        & $18.592\pm 0.0027$                            & {\it Gaia}                            \\
$J$                                        & 15.400$\pm0.060$                            & {\it 2MASS}                            \\
$H$                                        & 14.806$\pm0.067$                            & {\it 2MASS}                            \\
$K$                                       & 14.534$\pm0.061$                            & {\it 2MASS}                            \\
\noalign{\medskip}
{\it Derived parameters}                         & \multicolumn{1}{l}{} & \multicolumn{1}{l}{} \\
$M_{\text{C}}$ ($M_{\sun}$)                            & $0.253 \pm 0.010$                            & This work                            \\
$R_{\text{C}}$ ($R_{\sun}$)                          & $0.263 \pm 0.010$                            & This work                            \\
$T_{\text{eff},\text{C}}$ (K)                       & $3397^{+77}_{-63}$                            & This work

\end{tabular}%
}
\end{center}
\end{table}
In order to determine whether star B is a background star or physically associated with the planetary host star, we apply aperture photometry to the AO images. Saturation in the frames was corrected for by dividing the flux counts by the integration time for each image, in addition to taking the transmittance of the ND filter into account. However, because the potential companion is located in the halo of the host star in our observations, we have to deal with that first. 

\textcolor{black}{Because the asymmetric PSF could introduce systematic errors in the flux measurements if performed via radial-profile-subtraction of star A, we choose the following approach to estimate the flux ratio in the bands: The halo} of the host star is suppressed by applying a high-pass filter with a width of 4 FWHM. 
The filter not only suppresses flux from the host star, but also reduces flux from star B. \textcolor{black}{High-pass filtering introduces a flux loss of the companion, but the asymmetry in PSF has less impact on the companion's photometry since no specific shape is assumed for the targets radial profile.} Following \citet{2016ApJ...825...53H}, \textcolor{black}{the} loss in flux is estimated by injecting an artificial stellar signal representing the possible companion into the original image, at an angular distance similar to the true signal. We found that the high-pass filter reduces the flux from the injected star by approximately $5\%$. Taking this into account, we derive magnitude differences for the host star and star B of $\Delta H_B = 4.474\pm 0.092$~mag and \textcolor{black}{$\Delta K'_B = 4.270\pm 0.036$~mag}. 

After applying the high pass filter we \textcolor{black}{employed aperture photometry and then} fitted 2D Gaussians to estimate the location of the nearby companion for each band. We find angular separations of $0.389\pm 0.008$~arcsec in the $H$-band and $0.411\pm 0.015$~arcsec in the $K'$-band for the close-in companion. 

\textcolor{black}{As an additional consistency check for the photometric flux derivation we performed a photometry analysis in the $K'$-band on a radial-profile subtracted image. This revealed a magnitude difference of $\Delta K'_B = 4.256\pm 0.008$~mag, consistent with the above analysis, $\Delta K'_B = 4.270\pm 0.036$~mag. Unfortunately due to the asymmetric PSF in the $H$-band, we could not perform such an analysis there. In the following we will use the latter value, such that our flux estimates for the $H$- and $K'$-bands are derived in a consistent way.}

Contrast analysis and aperture photometry was not performed for the outer star (star C), which is therefore not displayed in the inset images in fig. \ref{fig:AO}. This is because it is far enough away to not cause blending effects in the light curve of the host star, and because it was at the very edge of the detector in the AO images, complicating the contrast analysis. Furthermore, a sufficient number of literature values of the magnitudes is already available for a thorough stellar analysis of star C. 

We derive fundamental parameters of star B and star C using {\tt{BASTA}}. We assume a distance and metallicity similar to the host star. For star B, we fit the $H$ magnitude computed using the magnitude difference $\Delta H$ from the AO analysis and the combined $H$ magnitude of the host star and star B from 2MASS. \textcolor{black}{An absolute value of the $K'$ magnitude has not been measured for the two stars. We therefore use only the $H$ band for extracting stellar parameters for star B\footnote{\textcolor{black}{Even though the $K'$ band of IRCS (1.95--2.30~$\mu$m) is similar to the $K$ band of 2MASS (1.95--2.36~$\mu$m), we wanted to keep the analysis to bands in which we could strictly compare. However, assuming $K'=K$ and repeating the analysis gave the same results.}}.} For star C, we fit the 2MASS $JHK$ magnitudes. The masses, radii and temperatures of the companions are reported in Table\,\ref{table:companion_parameters}. \textcolor{black}{We stress that the uncertainties on the derived parameters are internal to the BaSTI isochrones used.} We place the three stars in an HR diagram, see Fig. \ref{fig:HR}. Star A is a slightly evolved F8 star while Star B and C are both M-dwarfs.

For the host star and companion C the {\it{Gaia}} DR2 catalog provides parallaxes of $3.64\pm 0.05$~mas and $4.05\pm 0.27$~mas, respectively. These translate to line-of-sight distances of $275\pm 4$~pc and $247^{+18}_{-15}$~pc, consistent with the analysis of the isochrones and our assumption of physical association. Star B is not resolved in the {\it{Gaia}} data. The angular separation of star B and C translates into separations of $113\pm 2$~AU and $2467^{+177}_{-155}$~AU from K2-290, using the parallax of the host star. \textcolor{black}{The close proximity of star B to star A makes it likely that the two stars are indeed also physically associated and that B is at the same distance from us as A and C. To quantify this statement we calculated the probability of a chance alignment for A and B making use of the {\tt Besan\c{c}on} Galactic population model\footnote{\url{http://modele2016.obs-besancon.fr}} \citep{2003A&A...409..523R}. Using the default parameters\footnote{\textcolor{black}{Specified in appendix \ref{appendix_besancon}.}}, the model predicts 2413 background sources as bright or brighter than star B in a 1~deg$^2$ area surrounding star A. Scaling to an area just enclosing star A and B (i.e. with a radius of $\sim 0.4$ arcsec), the probability of chance alignment is  $<0.01\%$. Given this value we assume in the following that star B is physically associated with star A, and acknowledge that this association is based on a probability statement.}
This seems to also be the case for companion C, since it shares the same proper motion as the host star (see Tables\,\ref{table:host_star_parameters} and \ref{table:companion_parameters}). \textcolor{black}{In conclusion, K2-290 is most likely a member of a triple star system.}

\section{Planetary analysis}
\label{sec:planet}

In this section we test whether the photometric transits are a result of a false positive scenario, in particular component B being an eclipsing binary. We then describe the transit model as well as our RV model, and how we jointly fit these to extract system parameters from the data.

\subsection{False positive analysis}
\label{sec:false_pos}
\begin{figure}
\centering
\includegraphics[width=\columnwidth]{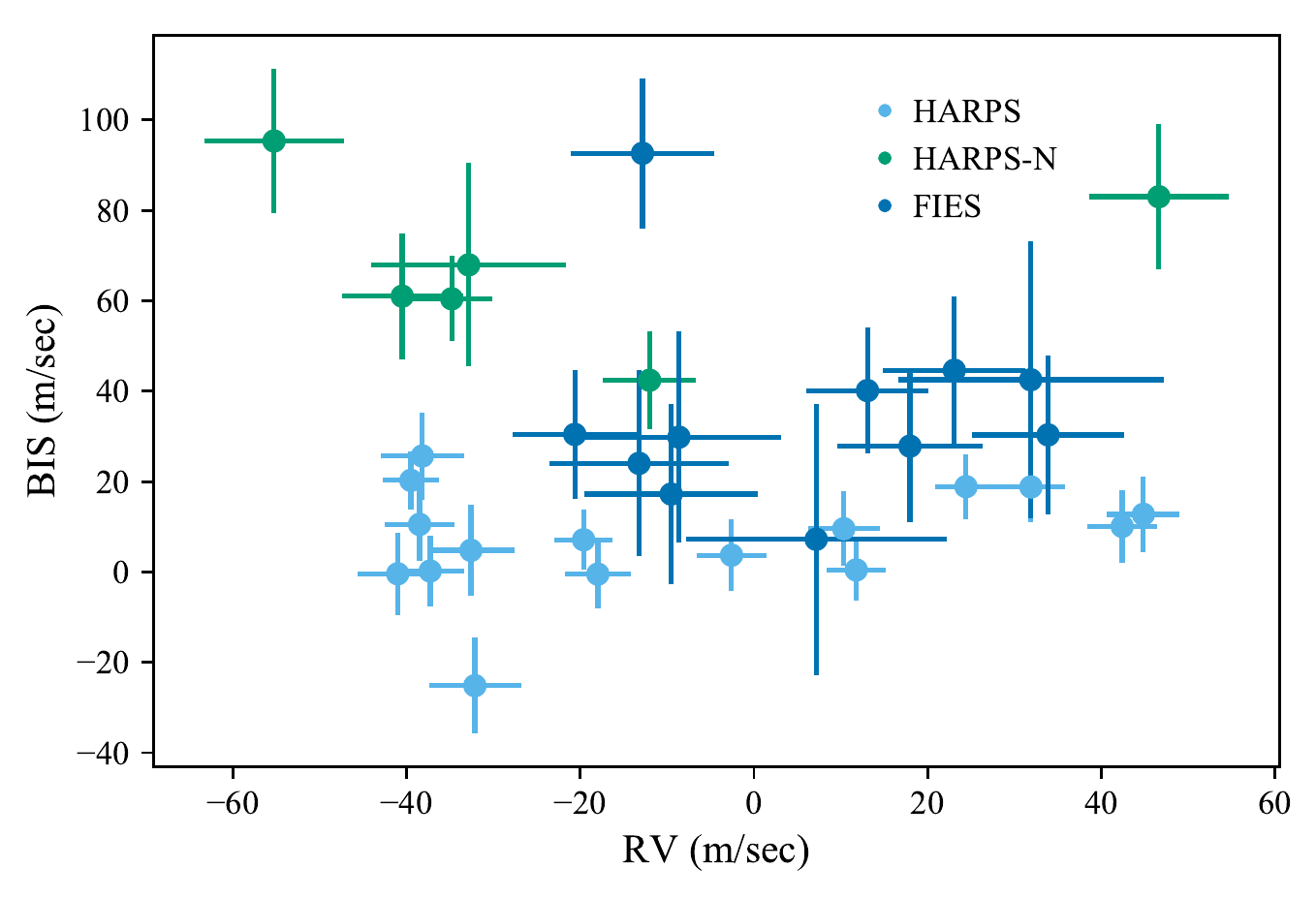}
\caption{Correlation between the CCF bisector inverse slopes and the radial velocities from the HARPS, HARPS-N and FIES spectrographs. The offsets for each spectrographs has been subtracted, with the best-fitting values found during the global modelling of the photometry and spectroscopy as described in Sec. \ref{subsub:global}.}
\label{fig:BIS}
\end{figure}
We test the scenarios in which the apparent transits do not originate from a planet occulting the host star, but instead from component B being a system of eclipsing binaries or being the host star of both planets. We do this because star B is not spatially resolved in the {\it{K2}} photometric lightcurve, due to its close proximity to the host star and pixel sizes of the spacecraft. The amount of flux received therefore also needs to be corrected, in order for the normalized transit to not appear too shallow. This is done by comparing the $H$\footnote{\textcolor{black}{As mentioned above,} an absolute value of the $K'$ magnitude has not been measured for the host star. We therefore use only the $H$ band for comparison between the two stars.}
magnitude for the companion to BaSTI isochrones, assuming the reddening, metallicity and age is the same as for the planetary host star. From this we can obtain {\it{Sloan}} $r$ and $g$ magnitudes of star B, which can be converted to a Kepler magnitude using the relation presented in \citet{Brown2011}. This analysis reveals that the close-by companion is $\sim 7.2$~mag fainter in the Kepler band-pass, corresponding to a flux contribution of $\sim 0.1-0.2 \%$ in the light curve. This indicates that the \textcolor{black}{large planet} must be orbiting the bright star, since star B is too faint and its light is too red to account for transits of the observed depths in the Kepler-band: Assuming the companion is totally eclipsed the blended transit depth will only be the afforementioned $\sim 0.1-0.2 \%$. This is too shallow to produce the deepest transits, which have depths of $0.5 \%$. If the smallest transit signals is due to the companion being an eclipsing binary diluting the signal, the transit depth of $0.03 \%$ would mean that $\sim 15-30 \%$ of the companion should be covered during transit. This would lead to a V-shaped transit, which is inconsistent with what we observe (see left panel of Fig. \ref{fig:lc}). Therefore, both planets are highly unlikely to be false positives. In the system analysis, the blending of the close companion is taken into account by subtracting its flux contribution from the photometric light curve.

Another analysis can be done by examining the asymmetry of the line profile, via investigating whether there is a correlation between the CCF bisector inverse slopes (BISs) and the RVs \citep[e.g.][]{Queloz2001}. Fig. \ref{fig:BIS} displays the BIS as a function of the RV data, showing no signs of correlation -- particularly if each instrument is considered separately. This suggests that the Doppler shifts of K2-290 are due to the orbital motion of the large planet, and not an astrophysical false positive.

\textcolor{black}{A third false positive check can be performed by comparing stellar parameters from the analysis of the host star described in Sec. \ref{sec:star} with transit observables extracted as described in the following sections. Assuming circular orbits, we calculate stellar densities $\rho_{\star,\text{circ}}$ of $0.51\pm0.20$~g~cm$^{-3}$ and $0.55\pm0.07$~g~cm$^{-3}$ using the best-fitting parameters for planet b and c from an analysis without a prior on the stellar spectrocopic density $\rho_{\star}$. With the exclusion of this prior, we assume that the best-fitting parameters of the transits are not strongly linked to the extracted stellar parameters. These densities agrees with the value from the stellar analysis of the host star $\rho_{\star}=0.485 \pm 0.07$~g~cm$^{-3}$. Using the values of $R_{B}$, $R_{C}$, $M_{B}$ and $M_{C}$ from the companion analysis in Sec. \ref{subsec:companions} we retrieve mean densities of the stars of $\rho_{B}=12.2\pm 2.2$~g~cm$^{-3}$ and $\rho_{C}=20.1\pm 2.4$~g~cm$^{-3}$. These do not agree with the values obtained from the transit parameters, and are therefore inconsistent with the planets orbiting either of the two M dwarf companions, further verifying that both planets orbit star A.}

\subsection{Transit model}
From the photometric data, each transit is isolated in a window spanning 15~hr on either side of the mid-transit time. The photometric uncertainty $\sigma_P$ is estimated as the standard deviation of the normalized out-of-transit data in these windows. The transits are normalized individually by including a quadratic polynomial fit of the data to the transit model during the parameter evaluation described in Sec.\,\ref{subsub:global}.
The transit lightcurve with a quadratic limb-darkening profile is modeled using {\tt{batman}} \citep{Kreidberg2015}, a Python package which calculates the lightcurve analytically based on the formalism of \citet{MandelAndAgol2002}. When modelling the light curve, the {\it Kepler} 29.4~min integration time is mimicked by integrating over 10 models which had been evaluated in a time interval of 29.4~min. The free parameters for each transiting planet are the orbital period $P_{k}$, the mid-transit time $T_{0k}$, the scaled planetary radius $R_{\text{p}k}/R_{\star}$, the scaled orbital distance $a_k/R_{\star}$, and its orbital inclination $i_k$. The index $k$ runs over planet b and c. For planet c, which influence we could identify in the RV data (see Sec.\,\ref{sec:spectroscopy}), \textcolor{black}{we both investigate a circular and eccentric solution (see Sec.\,\ref{planet_parameters}). In the latter, the orbital eccentricity $e$ and the argument of periastron $\omega$ are treated as free parameters}. For efficiency we step in $\sqrt{e}\cos{\omega}$ and $\sqrt{e}\sin{\omega}$ \citep{Ford2006,2011ApJ...726L..19A}. We find that we cannot sufficiently constrain the eccentricity of planet b, and we therefore assume the orbit of the small planet to be circular. \textcolor{black}{This is consistent with \citet{VanEylen2015} and \citet{VanEylen2019}, which show that near-zero eccentricity is likely for a small planet in a system with multiple transiting planets, and that the eccentricity distribution of such planets can be described by the positive half of a Gaussian distribution, which peaks at zero eccentricity and has a width of $\sigma = 0.083^{+0.015}_{-0.020}$.} Stellar limb darkening is modeled assuming a quadratic limb-darkening law with parameters $c_1$ and $c_2$. Finally we introduce an additional term $\sigma_{\text{K2}}$ in an attempt to capture any unacccounted photometric noise (e.g.\ caused by planetary spot crossing), similar to the jitter term often used in the RV work. This is added in quadrature to the photometric errors. With the 735 photometric measurements considered here, the log-likelihood for the photometry alone then becomes: 

\begin{equation}
\ln \mathcal{L}_P = -\dfrac{1}{2} \sum\limits_{i=1}^{735} \left( \ln \left( 2 \pi \left[ \sigma_i^2 + \sigma_{\text{K2}}^2 \right] \right) + \dfrac{\left[ P_i(O)-P_i(C) \right]^2}{\left[ \sigma_P^2 + \sigma_{\text{K2}}^2 \right]} \right)
\label{logL_P}
\end{equation}
where $P_i(O)$ and $P_i(C)$ are the observed and calculated values of the $i$'th photometric data point, $\sigma_P=0.000056$ is the internal measurement uncertainty estimated from the out-of-transit lightcurve and $\sigma_{\text{K2}}$ contains any additional photometric noise.

\subsection{Radial velocity model}
The radial velocity shifts of the host star due to the gravitational pull of the planets is modeled with a simple Keplerian model. Because we found no signs of planet b in the RV data (see Sec.\,\ref{sec:spectroscopy}), our RV model only includes planet c.
The additonal parameters needed are the RV semi-amplitude $K$ and RV offsets $\gamma$ as well as jitter terms $\sigma_{\text{jit}}$ for each spectrograph. The latter accounts for any stellar or instrumental noise not captured in the internal uncertainties and is added in quadrature. The log-likelihood for the 33 RV data points is: 

\small
\begin{equation}
\label{logL_RV}
\ln \mathcal{L}_{RV} = -\dfrac{1}{2} \sum\limits_{j=1}^{33} \Biggl( \ln \left( 2 \pi \left[ \sigma_j^2 + \sigma_{\text{jit}}^2 \right] \right) + \dfrac{\left[ RV_j(O)-RV_j(C)-\gamma \right]^2}{\left[ \sigma_j^2 + \sigma_{\text{jit}}^2 \right]} \Biggr)
\end{equation}
\normalsize
where $j$ indexes the 33 observations. $RV_j(O)$ and $RV_j(C)$ are the observed and calculated values of the $j$'th RV data point at time $t_j$, with the corresponding internal measurement uncertainty $\sigma_j$, while $\gamma$ and $\sigma_{\text{jit}}$ are the RV offset and jitter parameters, which differ for each spectrograph.

\subsection{Comparing models and data}
\label{subsub:global}
\begin{figure*}
\centering
\includegraphics[width=\columnwidth]{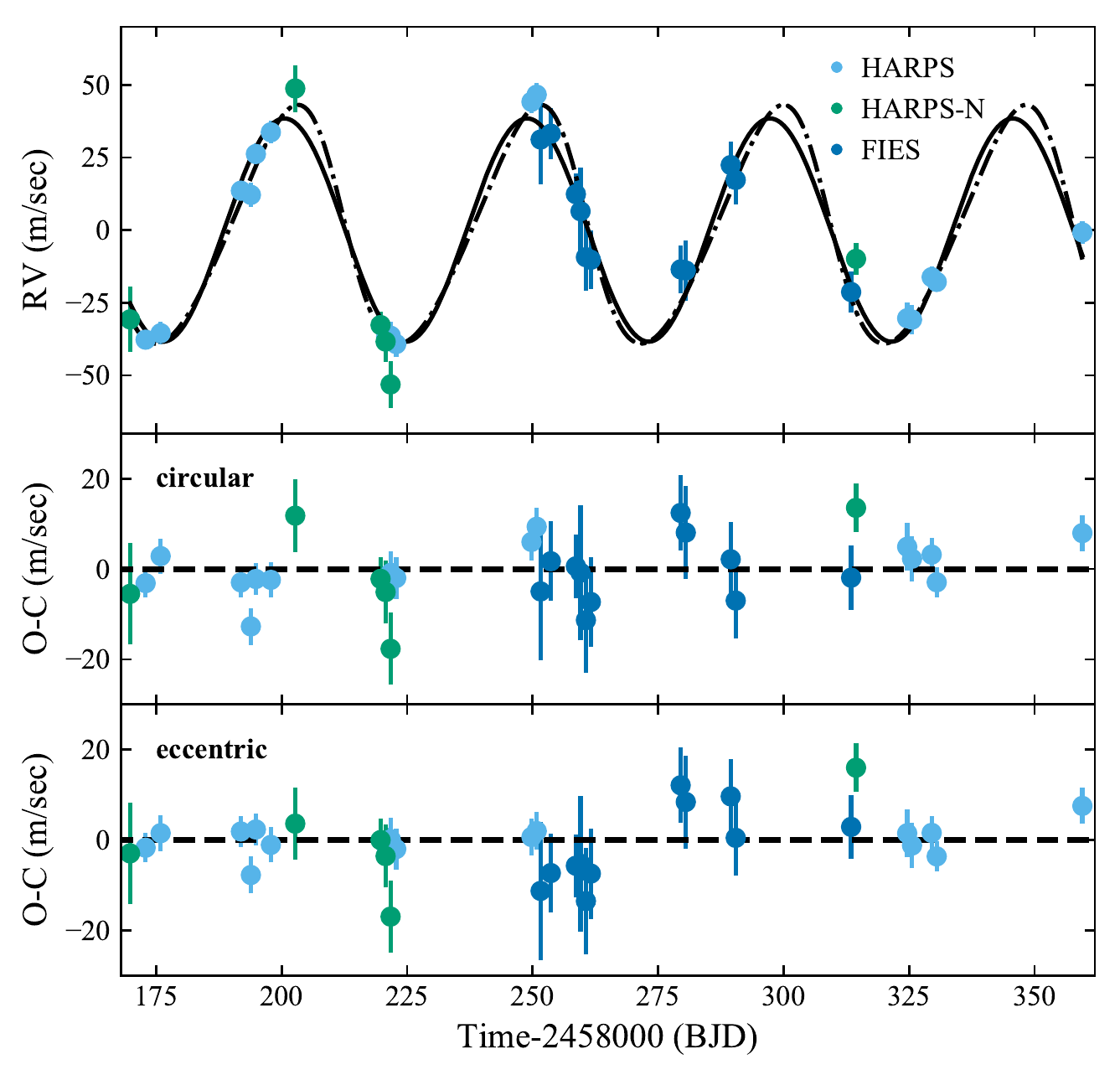}
\includegraphics[width=\columnwidth]{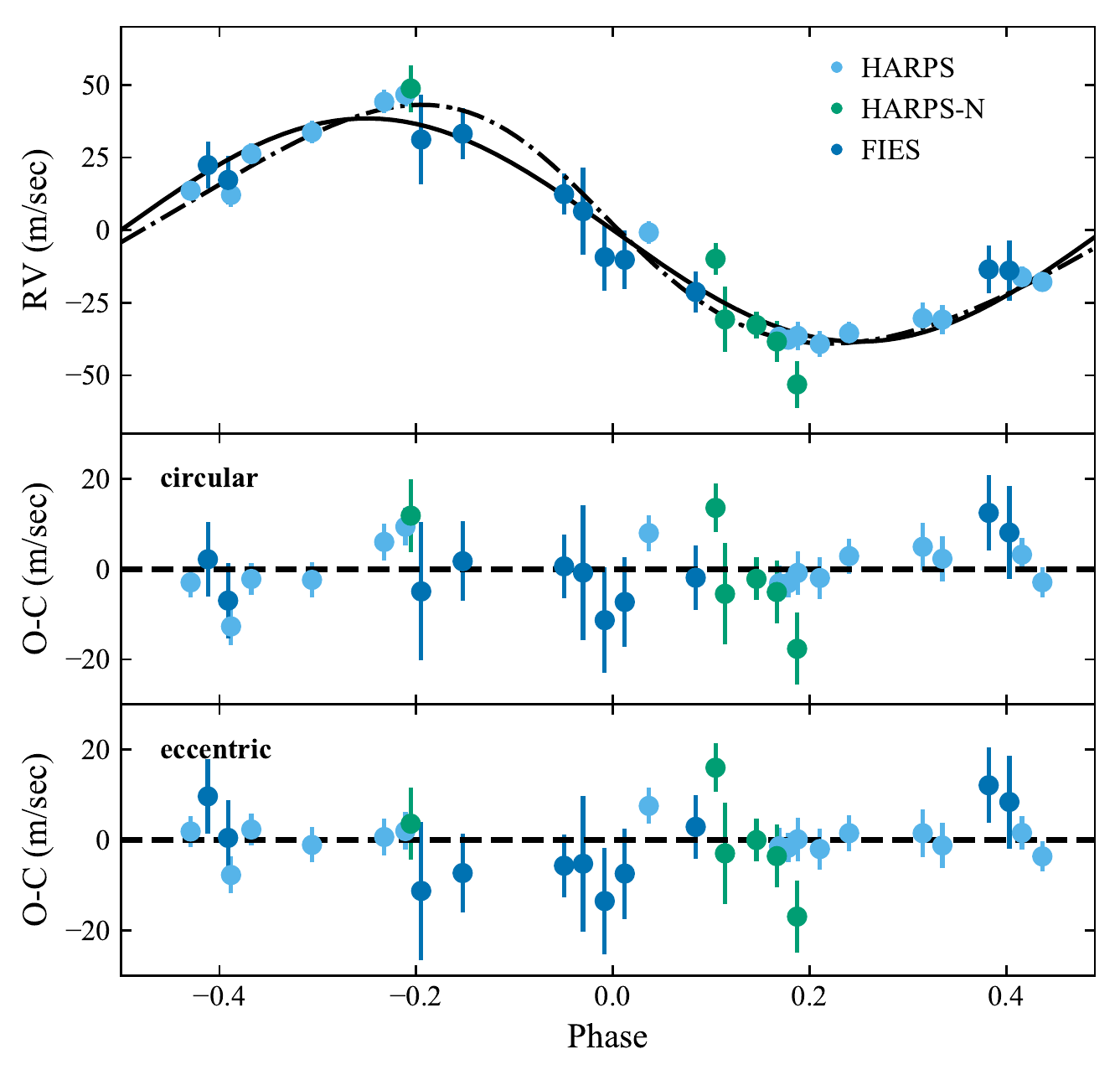}
\caption{RV measurements of K2-290c from the HARPS, HARPS-N and FIES spectrographs, together with the best-fitting \textcolor{black}{circular} model from the joint analysis of the photometry and spectroscopy (solid line) and the corresponding model for \textcolor{black}{an eccentric} orbit (dash-dotted line). Left: The RVs as a function of time. 
Right: The phasefolded RV. The bottom plots show the residuals between the observations and best-fitting model \textcolor{black}{for the circular and eccentric case. The eccentricity from the eccentric analysis is most likely overestimated and we therefore consider the circular model to be a better description of the data (see Sec. \ref{planet_parameters})}. The values of the corresponding parameters are displayed in Table \ref{table:system_parameters} \textcolor{black}{(Table \ref{table:system_parameters_ecc} for the eccentric case)}, and the data points are presented in Table\,\ref{table:RVs}.}
\label{fig:RV_planet_c}
\end{figure*}

\begin{figure*}
\centering
\includegraphics[width=\columnwidth]{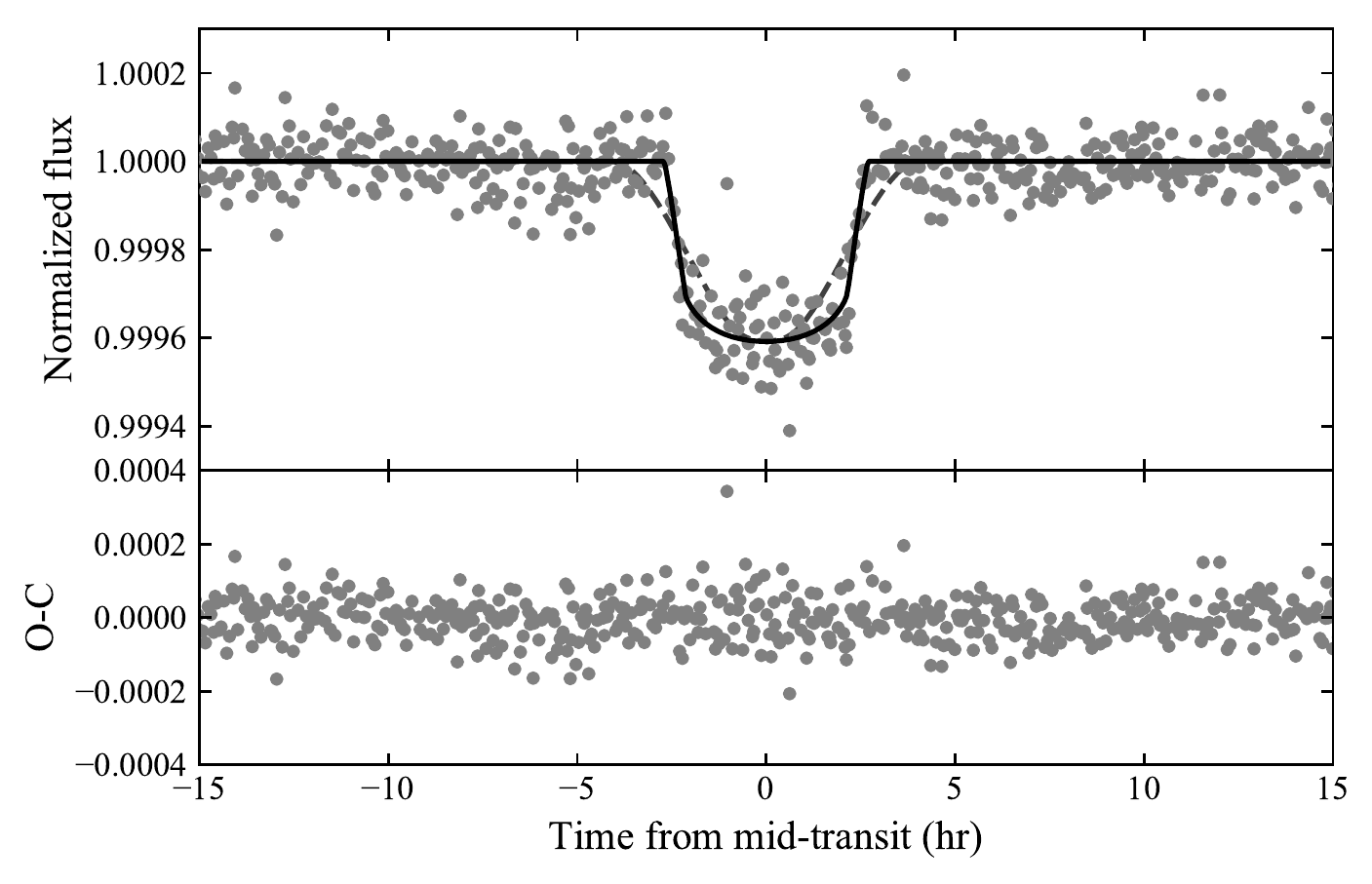}
\includegraphics[width=\columnwidth]{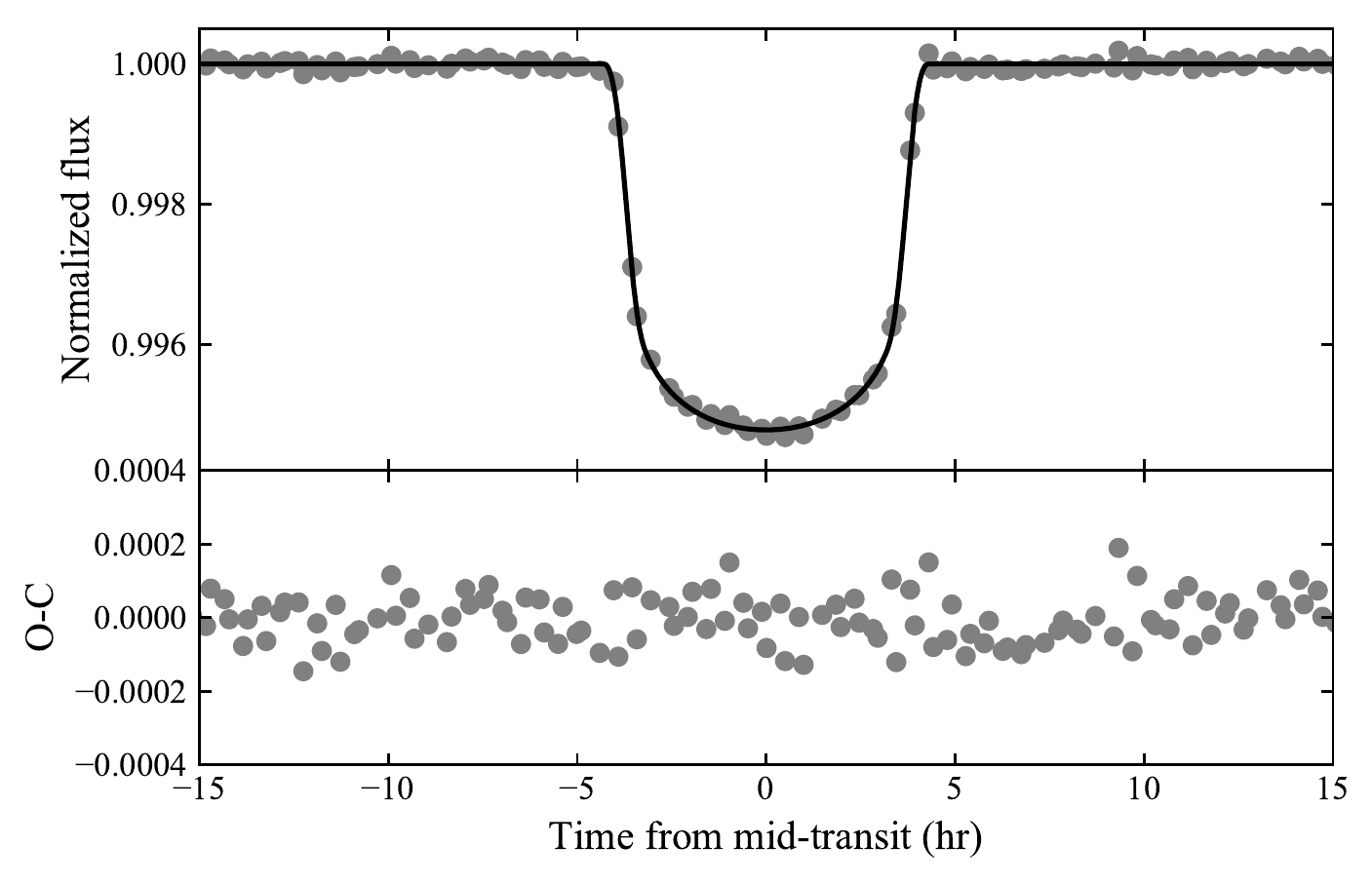}
\caption{Phasefolded transit light curves of K2-290b (left) and K2-290c (right) observed with K2, together with the best-fitting model from the joint analysis of the photometry and spectroscopy (solid line). The bottom plot shows the residuals. 
The values of the corresponding best-fit parameters are displayed in Table \ref{table:system_parameters}. 
The dashed line on the left plot indicates the modelled light curve in the case of the shallow transit signal being a false positive caused by star B. In order to reproduce the observed depth in the combined light of star A and B, this would require star B to be an eclipsing binary diluting star A with a transit depth of 15\% -- 30\%. This would lead to a very V-shaped transit, which is not what we observe. For the deep transit (right plot), even a total eclipse of star B is not sufficient to reproduce the signal.
}
\label{fig:lc}
\end{figure*}

To determine the parameters and their posterior distribution, we model the photometric and RV data together, fitting them jointly. 
In summary, the fitting parameters of the joint analysis are for each planet the orbital period $P$, the mid-transit time $T_{0}$, the scaled planetary radius $R_{\text{p}}/R_{\star}$, the scaled orbital distance $a/R_{\star}$, and its orbital inclination $i$. \textcolor{black}{For planet c, we also fit for the RV semi-amplitude $K$ and in addition we experiment with both a circular solution, as well as an eccentric analysis via the parametrization $\sqrt{e}\cos{\omega}$ and $\sqrt{e}\sin{\omega}$.} The fitting parameters connected to the star are the quadratic limb-darkening parameters $c_1$ and $c_2$. The fitting parameters for the instruments are the noise/jitter terms $\sigma$ and systemic RV velocities $\gamma$. 

For the limb-darkening coefficients we impose a Gaussian prior using the values $c_1 = 0.31$ and $c_2 = 0.30$ from an interpolation of the Kepler-band tables in \citet{ClaretandBloemen2011} obtained via \citet{EXOFAST}\footnote{\url{http://astroutils.astronomy.ohio-state.edu/exofast/limbdark.shtml}}, and with an uncertainty width of $0.1$. From the spectroscopic analysis we obtained a mean stellar density of the star $\rho_{\star} = 0.485\pm 0.07$\,g\,cm$^{-3}$. With a well-determined orbital period, we can use this information as an additional prior in our analysis, as photometric data also constrains the stellar density for particular orbital shapes and orientations \citep[see][and references therein]{VanEylen2015}. Therefore, we use this prior information and the transit photometry to support the $e$ and $\omega$ measurements from the RV data \textcolor{black}{when exploring the eccentric model}. 
The rest of the parameters are uniformly sampled.
The priors on $\rho_{\star}$, $c_1$ and $c_2$ have a log-likelihood $\ln \mathcal{L}_{\text{prior}}$.
The total log-likelihood is the sum of eq. \ref{logL_P}, \ref{logL_RV} and $\ln \mathcal{L}_{\text{prior}}$:
\begin{equation}
\ln \mathcal{L} = \ln \mathcal{L}_P + \ln \mathcal{L}_{RV} + \ln \mathcal{L}_{\text{prior}}.
\end{equation}
The posterior distribution of the fitting parameters are sampled using the MCMC Python package {\tt{emcee}} \citep{emcee}. We initalize 220 walkers near the maximum likelihood result, advancing them for 10000 steps and abandoning the 5000 first steps as the burnt-in sample, at which point the walkers have converged.

\subsection{Planet parameters}
\label{planet_parameters}

\begin{table*}
\begin{center}
\centering
\caption{System parameters for K2-290. {\bf{Notes:}} \textcolor{black}{*We both investigate a circular and eccentric solution. From the eccentric analysis we obtain $e=0.144^{+0.033}_{-0.032}$ and $\omega=70.0\pm 9.0$~deg. With $\omega$ close to $90$~deg the eccentricity from the eccentric analysis is most likely overestimated and we suspect that the circular model is a better description of the data (see Sec. \ref{planet_parameters}). Here we therefore only report the parameter values from the circular analysis, together with the one-sided $3\sigma$ upper limit on $e$ from the eccentric analysis. The complete set of parameter values of the eccentric solution is given in Table \ref{table:system_parameters_ecc}. $^{\dagger}$Upper limit (3$\sigma$) value obtained by including planet b in the RV analysis and allowing $e$ and $\omega$ for both planets to vary as well. The 1$\sigma$ results are given in the text in Sec. \ref{planet_parameters}. $^{\ddagger}$The values of the equlibrium temperatures assume a Bond albedo of 0 and no recirculation of heat. The errors only represent propagated internal errors.}}
\label{table:system_parameters}
\begin{tabular}{lcc}
\hline\hline
\noalign{\smallskip}
\multicolumn{3}{c}{Host star parameters (fixed)}                                                                               \\
\noalign{\smallskip}
\hline
Stellar mass $M_{\star}$ ($M_{\sun}$)                            & \multicolumn{2}{c}{$1.194^{+0.067}_{-0.077}$}                                       \\
Stellar radius $R_{\star}$ ($R_{\sun}$)                          & \multicolumn{2}{c}{$1.511^{+0.075}_{-0.072}$}                                       \\
Stellar density $\rho_{\star}$ (g cm$^{-3}$)                     & \multicolumn{2}{c}{$0.485^{+0.074}_{-0.064}$}                                       \\
Effective temperature $T_{\text{eff},\star}$ (K)                       & \multicolumn{2}{c}{$6302 \pm 120$}                                       \\
Surface gravity $\log g_{\star}$ (cgs)                                   & \multicolumn{2}{c}{$4.23 \pm 0.10$}                                       \\
Projected rotation speed $v\sin i_{\star}$ (km s$^{-1}$)                 & \multicolumn{2}{c}{$6.5 \pm 1.0$}                                       \\
Metallicity (Fe/H)                                               & \multicolumn{2}{c}{$-0.06 \pm 0.10$}                                       \\
Age (Gyr)                                                        & \multicolumn{2}{c}{$4.0^{+1.6}_{-0.8}$}                                       \\
\hline
\noalign{\smallskip}
Parameters from RV and transit MCMC analysis                         & Planet b & \textcolor{black}{Planet c (circular)*} \\
\noalign{\smallskip}
\hline
Quadratic limb darkening parameter $c_1$                             & \multicolumn{2}{c}{\textcolor{black}{$0.330\pm 0.044$}}                                       \\
Quadratic limb darkening parameter $c_2$                             & \multicolumn{2}{c}{$0.219\pm 0.067$}                                       \\
Noise term K2 $\sigma_{\text{K2}}$                             & \multicolumn{2}{c}{$0.0000209^{+0.0000044}_{-0.0000052}$}                                       \\
Jitter term FIES $\sigma_{\text{jit,FIES}}$ (m s$^{-1}$)         & \multicolumn{2}{c}{\textcolor{black}{$3.1_{-2.2}^{+3.5}$}}                                       \\
Jitter term HARPS $\sigma_{\text{jit,HARPS}}$ (m s$^{-1}$)       & \multicolumn{2}{c}{\textcolor{black}{$4.0_{-1.7}^{+1.8}$}}                                       \\
Jitter term HARPS-N $\sigma_{\text{jit,HARPS-N}}$ (m s$^{-1}$)   & \multicolumn{2}{c}{\textcolor{black}{$11.6_{-8.6}^{+5.3}$}}                                       \\
Systemic velocity FIES $\gamma_{\text{FIES}}$ (km s$^{-1}$)       & \multicolumn{2}{c}{\textcolor{black}{$19.6323_{-0.0030}^{+0.0031}$}}                                       \\
Systemic velocity HARPS $\gamma_{\text{HARPS}}$ (km s$^{-1}$)     & \multicolumn{2}{c}{\textcolor{black}{$19.7594\pm 0.0014$}}                                       \\
Systemic velocity HARPS-N $\gamma_{\text{HARPS-N}}$ (km s$^{-1}$) & \multicolumn{2}{c}{\textcolor{black}{$19.7590_{-0.0062}^{+0.0056}$}}                                       \\
Orbital period $P$ (days)                                        & $9.21165^{+0.00033}_{-0.00034}$     & \textcolor{black}{$48.36685^{+0.00041}_{-0.00040}$}                            \\
Time of midttransit $T_0$ (BJD)                             & $2457994.7725^{+0.0016}_{-0.0015}$       & \textcolor{black}{$2458019.17333\pm 0.00029$}                            \\
Scaled planetary radius $R_{\text{p}}/R_{\star}$                     & $0.01900\pm 0.00028$            & \textcolor{black}{$0.06848^{+0.00042}_{-0.00047}$}                            \\
Scaled orbital distance $a/R_{\star}$                                & $13.15^{+0.69}_{-0.66}$         & \textcolor{black}{$43.5\pm 1.2$}                            \\
Orbital inclination $i$ (deg)                                    & $88.14^{+0.62}_{-0.50}$             & \textcolor{black}{$89.37^{+0.08}_{-0.07}$}                            \\
RV semi-amplitude $K_{\star}$ (m s$^{-1}$)                       & < 6.6$^{\dagger}$                           & \textcolor{black}{$38.4 \pm 1.7$}                            \\
\hline
\noalign{\smallskip}
Derived parameters                                                   & \multicolumn{1}{l}{}         & \multicolumn{1}{l}{}         \\
\noalign{\smallskip}
\hline
Orbital eccentricity $e$                                             & 0 (adopted)                            & \textcolor{black}{0 (adopted, <0.241)}                            \\
Argument of periastron $\omega$ (deg)                            & 90 (adopted)                            & \textcolor{black}{90 (adopted)}                          \\
Impact parameter $b$                                             & \textcolor{black}{$0.438\pm 0.023$}                            & \textcolor{black}{$0.474\pm 0.012$}                            \\
Total transit duration $T_{14}$ (hr)                           & $4.96\pm 0.31$                            & \textcolor{black}{$8.14 \pm 0.26$}                            \\
Full transit duration \textcolor{black}{$T_{23}$} (hr)                   & $4.73\pm 0.40$                            & \textcolor{black}{$6.82\pm 0.24$}                            \\
Planetary mass $M_{\text{p}}$                   & < \textcolor{black}{21.1} $M_{\earth}$$^{\dagger}$                           & \textcolor{black}{$0.774\pm 0.047$ $M_{\text{J}}$}                           \\
Planetary radius $R_{\text{p}}$                 & $3.06\pm 0.16 R_{\earth}$                            & \textcolor{black}{$1.006\pm 0.050R_{\text{J}}$}                            \\
Planetary mean density $\rho_{\text{p}}$ (g cm$^{-3}$)                & \textcolor{black}{< 4.1$^{\dagger}$}                            & \textcolor{black}{$1.01\pm 0.16$}                            \\
semi-major axis a (AU)                                           & $0.0923\pm 0.0066$                            & \textcolor{black}{$0.305\pm 0.017$}                            \\
Equlibrium temperature $T_{\text{eq}}$ (K)                       & $1230\pm 38$\textcolor{black}{$^{\ddagger}$}                      & \textcolor{black}{$676\pm 16$$^{\ddagger}$}                            \\
\end{tabular}
\end{center}
\end{table*}

The parameter values corresponding to the median of the MCMC posterior distributions are reported in Table~\ref{table:system_parameters} together with their 1$\sigma$ uncertainties. The RVs and phasefolded RVs for planet c is shown in Fig. \ref{fig:RV_planet_c}, while the phasefolded lightcurves for planet b and c are displayed in Fig. \ref{fig:lc}. 

To account for any long-term trend from a possible long-period unseen companion, we could also allow for a linear drift of the RV signal, $\dot{\gamma}$. Including this in the analysis, and selecting BJD~$2458169.785818$ -- the time of the first RV observation -- as our zeropoint in defining $\dot{\gamma}$, we find a linear drift of $0.02\pm 0.02$~m~s$^{-1}$~d$^{-1}$. This shows that any possible RV trend is insignificant within $1\sigma$.
To further check whether we are justified in excluding a possible RV drift in our analysis, we compute the Bayesian Information Criterion (BIC). This is done for both an analysis including and excluding $\dot{\gamma}$. With 768 total RV and photometry measurements (as well as 3 priors), and 22 (23) model parameters excluding (including) the linear drift, we obtain a difference in BIC of 8. It favours the model {\it{excluding}} $\dot{\gamma}$, but we note that there are no significant differences in parameter values between the two models. The parameters values reported in Table~\ref{table:system_parameters} are for an analysis excluding the drift.

Because of the non-detection of K2-290b in the RV data (see Sec. \ref{sec:spectroscopy} and Fig. \ref{fig:LS}), it was not possible to confidently determine the mass of the planet. However, using the mass-radius relationship from \citet{Weiss2014}, the mass is estimated to be $\sim 7.6 M_{\earth}$. This is consistent with the smaller, close-in planet being a mini-Neptune. The mass translates into an RV semi-amplitude of $\sim$2-3~m~s$^{-1}$. Indeed, such signal would be hidden in the RVs, given the noise level of the data. \textcolor{black}{Doing an analysis which includes planet b in the RV fit and allows for varying $e$ and $\omega$ for the small planet, would indicate an RV semi-amplitude $K = 1.6^{+1.7}_{-1.1}$~m~s$^{-1}$, an eccentricity $e=0.119^{+0.201}_{-0.083}$ and a mass of $M_p = 5.8\pm 5.1 M_{\earth}$. Using the 3$\sigma$ result of this analysis, we obtain upper limits of $K < 6.6$~m~s$^{-1}$ and $M_p < 21.1 M_{\earth}$. We note that the phase coverage of the RVs of planet b is not ideal, with a large gap at phases $\sim$0.1-0.3. However, repeating the frequency analysis of Sec. \ref{sec:spectroscopy} but including noise-adjusted simulated data in this region with injected $K$-amplitudes up to 6.6~m~s$^{-1}$, still does not reveal signals above the $0.1\%$ FAP at the frequency of planet b (see Figure \ref{fig:LS_mockdata}).}

For K2-290c we find a mass of \textcolor{black}{$0.774\pm 0.047 M_J$} and a radius of \textcolor{black}{$1.006\pm 0.050 R_J$}. Together with its period of \textcolor{black}{$48.36685^{+0.00041}_{-0.00040}$}~days, this makes the planet a warm Jupiter. 

\begin{figure}
\centering
\includegraphics[width=\columnwidth]{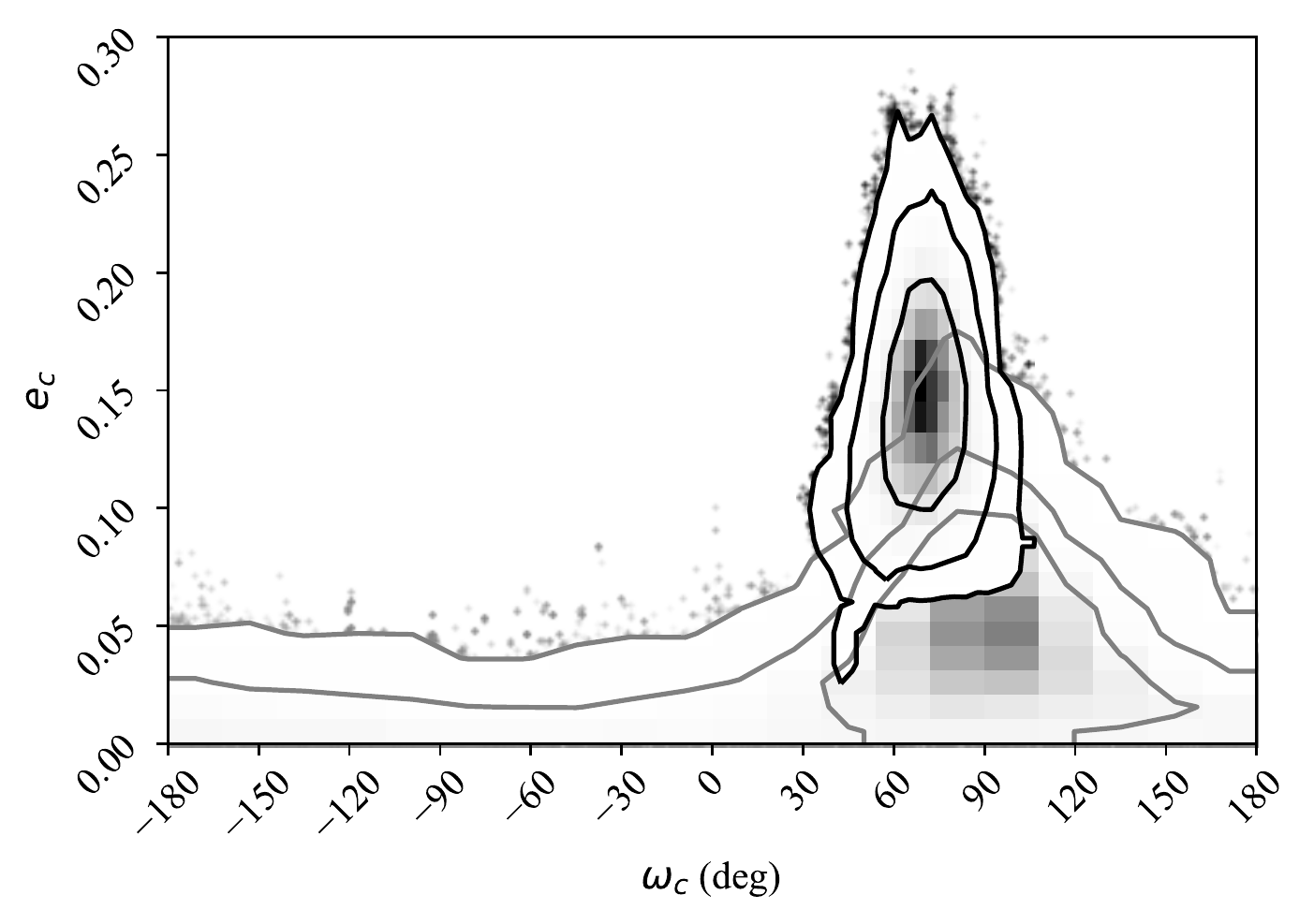}
\caption{\textcolor{black}{The 2D 68\%, 95\% and 99.7\% posterior distribution from the eccentric analysis described in Sec. \ref{subsub:global} (black) and from an analysis of RVs from a simulated circular orbit with added Gaussian noise corresponding to the real RV errors (grey). The analysis on the mock circular data allows for moderate eccentricities -- with its confidence limits overlapping near $\omega=90$ deg -- suggesting that with the data at hand we are not able to confirm a non circular orbit.}}
\label{fig:post_ecc_om_sim}
\end{figure}

\textcolor{black}{For the eccentric solution of planet c}, the posteriors of the eccentricity $e$ and argument of periastron $\omega$ is displayed in Fig. \ref{fig:post_ecc_om_sim} (black), and indicates that the planetary orbit is mildly eccentric with $e = 0.144^{+0.033}_{-0.032}$ \textcolor{black}{and $\omega=70.0\pm 9.0$~deg}. 
If we were not careful when removing the blended light from star B, \textcolor{black}{the eccentricity value} could be biased. \textcolor{black}{But, using no prior on the stellar density $\rho_{\star}$ -- and thereby essentially {\it{only}} obtaining information on the eccentricity from the RV data alone -- recovers an eccentricity $e = 0.130^{+0.037}_{-0.028}$, consistent with the previous analysis.}

\textcolor{black}{Doing the analysis for a circular orbit, and calculating the BIC of both the eccentric and circular fits, we can test whether we are justified in including $e$ and $\omega$ as two additional degrees of freedom. We obtain a difference in BIC of 16 in favour of the eccentric solution, suggesting that the eccentricity of planet c is well determined.}

\textcolor{black}{However, we note that $\omega=70.0\pm 9.0$~deg is close to $90$~deg. This warrants some further scrutiny as the RV method is better at constraining $\sqrt{e} \cos(\omega)$ than $\sqrt{e} \sin(\omega)$. Therefore larger confidence intervals for the eccentricity are allowed for orbital orientations near $-90$~deg or $+90$~deg than near $0$~deg or $\pm 180$~deg. Combined with an uneven phase coverage and the use of different instruments this could lead to an erroneous detection of a seemingly significant eccentricity \citep[e.g.][]{Laughlin2005,Albrecht_Wasp7}. We investigated this here by creating a mock data set, where we used the parameters of the circular solution from Table \ref{table:system_parameters}. With this circular model we now created for each of the original RV time stamp an RV "data point" adding random Gaussian noise corresponding to the RV uncertainties of the original data. Finally we run our analysis on this simulated data set just as we did for the real measurements. We repeated this experiment several times, using different seeds for the Gaussian noise. A typical example of the posterior of $e$ and $\omega$ for the simulated circular data is shown in Fig. \ref{fig:post_ecc_om_sim} (grey), together with the posterior from the eccentric analysis on the real data (black). We find that the uncertainty intervals for $e$ are largest around $\omega=+90$~deg, and indeed the 2D confidence intervals between the mock and real data do overlap. This suggests that the eccentricity we find from the eccentric analysis of the real data is suspicious and should serve as an upper bound on the eccentricity only. We therefore adobt the circular solution, which parameters are reported in Table \ref{table:system_parameters}, and note that from the eccentric analysis the one-sided $3\sigma$ upper limit on the eccentricity is $e<0.241$. Nonetheless, varying $e$ and $\omega$ only reveals minor changes in the rest of the system parameters, with almost all being within $1\sigma$ of the circular values (see Table \ref{table:system_parameters_ecc}).}

\section{Discussion}
\label{sec:discussion}
\subsection{Properties and composition of the planets}

Planet b is exposed to intense radiation from the host star. With a distance to the star of $0.0923\pm 0.0066$~AU or $13.15^{+0.69}_{-0.66} R_{\star}$, it receives an incident flux of $\sim 400 F_{\earth}$. This puts it  outside of the super-Earth desert \citep{Lundkvist2016}. It also resides above the radius valley \citep{Fulton2017, VanEylen2018}, suggesting that the planet is not undergoing photo-evaporation of its outer envelope.

Given the relatively low incident flux of planet c of $\sim 0.6 \cdot 10^8$ erg~s$^{-1}$~cm$^{-2}$, the planet lies below the threshold of $2 \cdot 10^8$ erg~s$^{-1}$~cm$^{-2}$, where irradiation might inflate it \citep{Demory2011}. The planetary radius may therefore be directly compared to the models presented by \citet{Fortney2007}, revealing a mass of the planetary core of about $\sim 25-50 M_{\earth}$\footnote{From the extended \href{ucolick.org/~jfortney/models/tbl_Evpcore_50a_point1AU.dat}{webtable} of \citet{Fortney2007}.}. \textcolor{black}{However, in these models all solids are assumed to be located in the core. The models of \citet{Thorngren2016} allow for metal enrichment and for solid materials to be located in the planet's gaseous envelope. Using these semi-empirical models, we retrieve a planetary bulk metallicity $Z=0.133\pm 0.036$ and a heavy elements mass of $49.5\pm 6.4 M_{\earth}$, with $10 M_{\earth}$ distributed inside the core and the remaining mixed in the envelope.}

\subsection{Formation}
We find that the orbit of the warm Jupiter K2-290c has an eccentricity \textcolor{black}{$e < 0.241$, and the existing RV data are compatible with a circular orbit}. This is consistent with the picture presented in \citet{2013ApJ...767L..24D}, where warm Jupiters with low eccentricities orbit metal-poor stars ([Fe/H]\,$=-0.06\pm0.1$). The orbital eccentricity is too small for the planet to be a proto hot Jupiter undergoing migration through tidal friction \citep[Fig. 4]{Dawson2018}. This does not rule out high-eccentricity migration through secular gravitational interactions, causing the planetary eccentricity to undergo oscillations excited by a nearby mutually inclined third body \citep{Petrovich2016}.
For this to happen, a solar-mass perturber needs to be within a distance of $\sim$~30~AU, and a Jupiter-mass perturber within $\sim$~3~AU \citep{Dong2014}, for a warm Jupiter $0.2$\,AU away. With a projected distance of $113 \pm 2$ AU even star B -- the closest companion -- is too far away. Neither from the AO images nor the transit light curve do we find evidence for an additional close-by companion.
Furthermore, it seems unlikely that the warm Jupiter and mini-Neptune would remain coplanar following these orbital perturbations, which is likely to produce higher mutual inclinations \citep{2018MNRAS.478..197P}. However seeing both planets in transit do not necessarily guarantee coplanar orbits, as we might have observed them along the line of nodes. 
It should also be noted, that even though the distances between the host star and its two stellar companions are in agreement with the outcome of simulated high-eccentricity migration of Jupiters in triple star systems in \citet{Hamers2017}, these simulations fail to produce warm Jupiters in any significant number.

With an eccentricity $<0.4$ and with the presence of its mini-Neptune sibling, K2-290c fits the picture presented in \citet{Huang2016}: low eccentricity warm Jupiter systems have inner low-mass companions with low mutual inclinations. They argue that this suggests that warm Jupiter might originate from two different formation mechanisms: 1) high-eccentricity systems ($e>0.4$) are formed through high-eccentricity migration and 2) low-eccentricity systems form {\it{in situ}}, since disk migration would clear out any companions in the warm Jupiter neighbourhood. The latter is consistent with the core mass of \textcolor{black}{$\sim 10-50 M_{\earth}$}, which is sufficient for the run-away accretion phase of {\it{in situ}} gas giant formation at distances of 0.1-1.0 AU from the host star \citep{Rafikov2006}. However, as noted in \citet{Dawson2018}, disk migration should not be ruled out as the origin channel of the warm Jupiter in these kind of systems, as the migration of the giant planet might have occured before the {\it{in situ}} creation of its small sibling.
This suggests that K2-290c originate from either {\it{in situ}} formation or disk migration.

A way to further test the origin of K2-290c would be to measure the system's spin-orbit angle. Here, alignment would point towards the system having been dynamically stable and formed {\it{in situ}} or through disk migration, while misalignment would suggest early instabilities and migration. The spin-orbit angle can be measured through the Rossiter-McLaughlin (RM) effect (\citealp{Rossiter}; \citealp{McLaughlin}), of which K2-290 is an excellent target: From the values of $v\sin i_{\star} \sim 6.5 \pm 1.0$ km~s$^{-1}$ and $R_p/R_{\star} \sim 0.07$, we expect an amplitude of the RM signal of about $\sim$19 m~s$^{-1}$, taking limb darkening and the eccentricity into account. The host star is bright ($V=11.11$), allowing for a high SNR and small RV errors, which makes the RM effect easily detectable with high resolution fiber-fed stabilized spectrographs. In addition, with an impact parameter of \textcolor{black}{$\sim 0.5$}, there should be no degeneracy between the spin-orbit angle and $v\sin i_{\star}$.

\section*{Acknowledgements}
\textcolor{black}{We are very grateful for the helpful comments and suggestions from the anonymous referee, which improved the quality of the paper.}
We \textcolor{black}{also} give our sincerest thanks to amateur astronomers Phil Evans and Chris Stockdale for their effort in obtaining ground-based photometry during transit of K2-290c.
MH, ABJ and SA acknowledge the support from the Danish Council for Independent Research through the DFF Sapere Aude Starting Grant No. 4181-00487B, and the Stellar Astrophysics Centre which funding is provided by The Danish National Research Foundation (Grant agreement no.: DNRF106).
\textcolor{black}{This project has received funding from the European Union's Horizon 2020 research and innovation programme under grant agreement No 730890. This material reflects only the authors views and the Commission is not liable for any use that may be made of the information contained therein.}
\textcolor{black}{The work was also} supported by Japan Society for Promotion of Science (JSPS) KAKENHI Grant Number JP16K17660 and partly supported by JSPS KAKENHI Grant Number JP18H01265.
ME acknowledges the support of the DFG priority program SPP 1992 "Exploring the Diversity of Extrasolar Planets" (HA 3279/12-1).
This paper includes data collected by the K2 mission. Funding for the K2 mission is provided by the NASA Science Mission directorate. Some of the data presented in this paper were obtained from the Mikulski Archive for Space Telescopes (MAST). STScI is operated by the Association of Universities for Research in Astronomy, Inc., under NASA contract NAS5-26555.
The radial velocity observations were made with 1) the Nordic Optical Telescope (NOT), operated by the Nordic Optical Telescope Scientific Association at the Observatorio del Roque de los Muchachos, La Palma, Spain, of the Instituto de Astrofisica de Canarias as part of the Nordic and OPTICON programmes 57-015 and 2018A/044, 2) the European Organisation for Astronomical Research in the Southern Hemisphere under ESO programmes 0100.C-0808 and 0101.C-0829 and 3) the Italian Telescopio Nazionale Galileo (TNG) operated on the island of La Palma by the Fundaci{\'o}n Galileo Galilei of the INAF (Istituto Nazionale di Astrofisica) at the Spanish Observatorio del Roque de los Muchachos of the Instituto de Astrofisica de Canarias as part of the Spanish and TAC programmes CAT17B\_99, CAT18A\_130, and A37TAC\_37.
The AO imaging was based on data collected at Subaru Telescope, which is operated by the National Astronomical Observatory of Japan as part of the programme S18A-089. The authors wish to recognize and acknowledge the very significant cultural role and reverence that the summit of Maunakea has always had within the indigenous Hawaiian community. We are most fortunate to have the opportunity to conduct observations from this mountain.
This work uses results from the European Space Agency (ESA) space mission Gaia. Gaia data are being processed by the Gaia Data Processing and Analysis Consortium (DPAC). Funding for the DPAC is provided by national institutions, in particular the institutions participating in the Gaia MultiLateral Agreement (MLA). The Gaia mission website is \url{https://cosmos.esa.int/gaia}. The Gaia archive website is \url{https://archives.esac.esa.int/gaia}.
\textcolor{black}{IRAF is distributed by the National Optical Astronomy Observatory, which is operated by the Association of Universities for Research in Astronomy (AURA) under a cooperative agreement with the National Science Foundation.}


\bibliographystyle{mnras}
\bibliography{Bibliography_vs2} 

\begin{thebibliography}{}
\makeatletter
\relax
\def\mn@urlcharsother{\let\do\@makeother \do\$\do\&\do\#\do\^\do\_\do\%\do\~}
\def\mn@doi{\begingroup\mn@urlcharsother \@ifnextchar [ {\mn@doi@}
  {\mn@doi@[]}}
\def\mn@doi@[#1]#2{\def\@tempa{#1}\ifx\@tempa\@empty \href
  {http://dx.doi.org/#2} {doi:#2}\else \href {http://dx.doi.org/#2} {#1}\fi
  \endgroup}
\def\mn@eprint#1#2{\mn@eprint@#1:#2::\@nil}
\def\mn@eprint@arXiv#1{\href {http://arxiv.org/abs/#1} {{\tt arXiv:#1}}}
\def\mn@eprint@dblp#1{\href {http://dblp.uni-trier.de/rec/bibtex/#1.xml}
  {dblp:#1}}
\def\mn@eprint@#1:#2:#3:#4\@nil{\def\@tempa {#1}\def\@tempb {#2}\def\@tempc
  {#3}\ifx \@tempc \@empty \let \@tempc \@tempb \let \@tempb \@tempa \fi \ifx
  \@tempb \@empty \def\@tempb {arXiv}\fi \@ifundefined
  {mn@eprint@\@tempb}{\@tempb:\@tempc}{\expandafter \expandafter \csname
  mn@eprint@\@tempb\endcsname \expandafter{\@tempc}}}

\bibitem[\protect\citeauthoryear{{Albrecht}, {Winn}, {Butler}, {Crane},
  {Shectman}, {Thompson}, {Hirano}  \& {Wittenmyer}}{{Albrecht}
  et~al.}{2012a}]{Albrecht_Wasp7}
{Albrecht} S.,  {Winn} J.~N.,  {Butler} R.~P.,  {Crane} J.~D.,  {Shectman}
  S.~A.,  {Thompson} I.~B.,  {Hirano} T.,   {Wittenmyer} R.~A.,  2012a, \mn@doi
  [\apj] {10.1088/0004-637X/744/2/189}, \href
  {http://adsabs.harvard.edu/abs/2012ApJ...744..189A} {744, 189}

\bibitem[\protect\citeauthoryear{{Albrecht} et~al.,}{{Albrecht}
  et~al.}{2012b}]{Albrecht2012}
{Albrecht} S.,  et~al., 2012b, \mn@doi [\apj] {10.1088/0004-637X/757/1/18},
  \href {http://adsabs.harvard.edu/abs/2012ApJ...757...18A} {757, 18}

\bibitem[\protect\citeauthoryear{{Anderson} et~al.,}{{Anderson}
  et~al.}{2011}]{2011ApJ...726L..19A}
{Anderson} D.~R.,  et~al., 2011, \mn@doi [\apjl] {10.1088/2041-8205/726/2/L19},
  \href {http://adsabs.harvard.edu/abs/2011ApJ...726L..19A} {726, L19}

\bibitem[\protect\citeauthoryear{{Anglada-Escud{\'e}}
  et~al.,}{{Anglada-Escud{\'e}} et~al.}{2012}]{Anglada2012}
{Anglada-Escud{\'e}} G.,  et~al., 2012, \mn@doi [\apjl]
  {10.1088/2041-8205/751/1/L16}, \href
  {http://adsabs.harvard.edu/abs/2012ApJ...751L..16A} {751, L16}

\bibitem[\protect\citeauthoryear{{Batygin}, {Bodenheimer}  \&
  {Laughlin}}{{Batygin} et~al.}{2016}]{Batygin2016}
{Batygin} K.,  {Bodenheimer} P.~H.,   {Laughlin} G.~P.,  2016, \mn@doi [\apj]
  {10.3847/0004-637X/829/2/114}, \href
  {http://adsabs.harvard.edu/abs/2016ApJ...829..114B} {829, 114}

\bibitem[\protect\citeauthoryear{{Blanco-Cuaresma}, {Soubiran}, {Heiter}  \&
  {Jofr{\'e}}}{{Blanco-Cuaresma} et~al.}{2014}]{2014A&A...569A.111B}
{Blanco-Cuaresma} S.,  {Soubiran} C.,  {Heiter} U.,   {Jofr{\'e}} P.,  2014,
  \mn@doi [\aap] {10.1051/0004-6361/201423945}, \href
  {http://adsabs.harvard.edu/abs/2014A%26A...569A.111B} {569, A111}

\bibitem[\protect\citeauthoryear{{Bonomo} et~al.,}{{Bonomo}
  et~al.}{2017}]{Bonomo2017}
{Bonomo} A.~S.,  et~al., 2017, \mn@doi [\aap] {10.1051/0004-6361/201629882},
  \href {http://esoads.eso.org/abs/2017A%26A...602A.107B} {602, A107}

\bibitem[\protect\citeauthoryear{{Borucki} et~al.,}{{Borucki}
  et~al.}{2010}]{Kepler}
{Borucki} W.~J.,  et~al., 2010, \mn@doi [Science] {10.1126/science.1185402},
  \href {http://adsabs.harvard.edu/abs/2010Sci...327..977B} {327, 977}

\bibitem[\protect\citeauthoryear{{Brown}, {Latham}, {Everett}  \&
  {Esquerdo}}{{Brown} et~al.}{2011}]{Brown2011}
{Brown} T.~M.,  {Latham} D.~W.,  {Everett} M.~E.,   {Esquerdo} G.~A.,  2011,
  \mn@doi [\aj] {10.1088/0004-6256/142/4/112}, \href
  {http://adsabs.harvard.edu/abs/2011AJ....142..112B} {142, 112}

\bibitem[\protect\citeauthoryear{{Campante} et~al.,}{{Campante}
  et~al.}{2015}]{Campante2015}
{Campante} T.~L.,  et~al., 2015, \mn@doi [\apj] {10.1088/0004-637X/799/2/170},
  \href {http://adsabs.harvard.edu/abs/2015ApJ...799..170C} {799, 170}

\bibitem[\protect\citeauthoryear{{Casagrande} \& {VandenBerg}}{{Casagrande} \&
  {VandenBerg}}{2014}]{CasagrandeVanderberg2014}
{Casagrande} L.,  {VandenBerg} D.~A.,  2014, \mn@doi [\mnras]
  {10.1093/mnras/stu1476}, \href
  {http://adsabs.harvard.edu/abs/2014MNRAS.444..392C} {444, 392}

\bibitem[\protect\citeauthoryear{{Casagrande} \& {VandenBerg}}{{Casagrande} \&
  {VandenBerg}}{2018}]{CasagrandeVanderberg2018}
{Casagrande} L.,  {VandenBerg} D.~A.,  2018, \mn@doi [\mnras]
  {10.1093/mnras/sty149}, \href
  {http://adsabs.harvard.edu/abs/2018MNRAS.475.5023C} {475, 5023}

\bibitem[\protect\citeauthoryear{{Claret} \& {Bloemen}}{{Claret} \&
  {Bloemen}}{2011}]{ClaretandBloemen2011}
{Claret} A.,  {Bloemen} S.,  2011, \mn@doi [\aap]
  {10.1051/0004-6361/201116451}, \href
  {http://adsabs.harvard.edu/abs/2011A%26A...529A..75C} {529, A75}

\bibitem[\protect\citeauthoryear{{Cosentino} et~al.,}{{Cosentino}
  et~al.}{2012}]{HARPSN}
{Cosentino} R.,  et~al., 2012, in Ground-based and Airborne Instrumentation for
  Astronomy IV. p. 84461V, \mn@doi{10.1117/12.925738}

\bibitem[\protect\citeauthoryear{{Cutri} et~al.,}{{Cutri} et~al.}{2003}]{2MASS}
{Cutri} R.~M.,  et~al., 2003, VizieR Online Data Catalog, \href
  {http://adsabs.harvard.edu/abs/2003yCat.2246....0C} {2246}

\bibitem[\protect\citeauthoryear{{Dai} et~al.,}{{Dai} et~al.}{2017}]{Dai2017}
{Dai} F.,  et~al., 2017, \mn@doi [\aj] {10.3847/1538-3881/aa9065}, \href
  {http://adsabs.harvard.edu/abs/2017AJ....154..226D} {154, 226}

\bibitem[\protect\citeauthoryear{{Dawson} \& {Johnson}}{{Dawson} \&
  {Johnson}}{2018}]{Dawson2018}
{Dawson} R.~I.,  {Johnson} J.~A.,  2018, preprint, \href
  {http://adsabs.harvard.edu/abs/2018arXiv180106117D} {} (\mn@eprint {arXiv}
  {1801.06117})

\bibitem[\protect\citeauthoryear{{Dawson} \& {Murray-Clay}}{{Dawson} \&
  {Murray-Clay}}{2013}]{2013ApJ...767L..24D}
{Dawson} R.~I.,  {Murray-Clay} R.~A.,  2013, \mn@doi [\apjl]
  {10.1088/2041-8205/767/2/L24}, \href
  {http://adsabs.harvard.edu/abs/2013ApJ...767L..24D} {767, L24}

\bibitem[\protect\citeauthoryear{{Demory} \& {Seager}}{{Demory} \&
  {Seager}}{2011}]{Demory2011}
{Demory} B.-O.,  {Seager} S.,  2011, \mn@doi [\apjs]
  {10.1088/0067-0049/197/1/12}, \href
  {http://adsabs.harvard.edu/abs/2011ApJS..197...12D} {197, 12}

\bibitem[\protect\citeauthoryear{{Dong}, {Katz}  \& {Socrates}}{{Dong}
  et~al.}{2014}]{Dong2014}
{Dong} S.,  {Katz} B.,   {Socrates} A.,  2014, \mn@doi [\apjl]
  {10.1088/2041-8205/781/1/L5}, \href
  {http://adsabs.harvard.edu/abs/2014ApJ...781L...5D} {781, L5}

\bibitem[\protect\citeauthoryear{{Eastman}, {Gaudi}  \& {Agol}}{{Eastman}
  et~al.}{2013}]{EXOFAST}
{Eastman} J.,  {Gaudi} B.~S.,   {Agol} E.,  2013, \mn@doi [\pasp]
  {10.1086/669497}, \href {http://adsabs.harvard.edu/abs/2013PASP..125...83E}
  {125, 83}

\bibitem[\protect\citeauthoryear{{Feroz} \& {Hobson}}{{Feroz} \&
  {Hobson}}{2014}]{Feroz2014}
{Feroz} F.,  {Hobson} M.~P.,  2014, \mn@doi [\mnras] {10.1093/mnras/stt2148},
  \href {http://adsabs.harvard.edu/abs/2014MNRAS.437.3540F} {437, 3540}

\bibitem[\protect\citeauthoryear{{Ford}}{{Ford}}{2006}]{Ford2006}
{Ford} E.~B.,  2006, \mn@doi [\apj] {10.1086/500802}, \href
  {http://adsabs.harvard.edu/abs/2006ApJ...642..505F} {642, 505}

\bibitem[\protect\citeauthoryear{{Foreman-Mackey}, {Hogg}, {Lang}  \&
  {Goodman}}{{Foreman-Mackey} et~al.}{2013}]{emcee}
{Foreman-Mackey} D.,  {Hogg} D.~W.,  {Lang} D.,   {Goodman} J.,  2013, \mn@doi
  [\pasp] {10.1086/670067}, \href
  {http://adsabs.harvard.edu/abs/2013PASP..125..306F} {125, 306}

\bibitem[\protect\citeauthoryear{{Fortney}, {Marley}  \& {Barnes}}{{Fortney}
  et~al.}{2007}]{Fortney2007}
{Fortney} J.~J.,  {Marley} M.~S.,   {Barnes} J.~W.,  2007, \mn@doi [\apj]
  {10.1086/512120}, \href {http://adsabs.harvard.edu/abs/2007ApJ...659.1661F}
  {659, 1661}

\bibitem[\protect\citeauthoryear{{Fulton} et~al.,}{{Fulton}
  et~al.}{2017}]{Fulton2017}
{Fulton} B.~J.,  et~al., 2017, \mn@doi [\aj] {10.3847/1538-3881/aa80eb}, \href
  {http://adsabs.harvard.edu/abs/2017AJ....154..109F} {154, 109}

\bibitem[\protect\citeauthoryear{{Gaia Collaboration}, {Brown}, {Vallenari},
  {Prusti}, {de Bruijne}, {Babusiaux}  \& {Bailer-Jones}}{{Gaia Collaboration}
  et~al.}{2018}]{Gaia2018}
{Gaia Collaboration} {Brown} A.~G.~A.,  {Vallenari} A.,  {Prusti} T.,  {de
  Bruijne} J.~H.~J.,  {Babusiaux} C.,   {Bailer-Jones} C.~A.~L.,  2018,
  preprint, \href {http://adsabs.harvard.edu/abs/2018arXiv180409365G} {}
  (\mn@eprint {arXiv} {1804.09365})

\bibitem[\protect\citeauthoryear{{Gandolfi} et~al.,}{{Gandolfi}
  et~al.}{2013}]{Gandolfi2013}
{Gandolfi} D.,  et~al., 2013, \mn@doi [\aap] {10.1051/0004-6361/201321901},
  \href {https://ui.adsabs.harvard.edu/#abs/2013A&A...557A..74G} {557, A74}

\bibitem[\protect\citeauthoryear{{Green} et~al.,}{{Green}
  et~al.}{2018}]{Green2018}
{Green} G.~M.,  et~al., 2018, \mn@doi [\mnras] {10.1093/mnras/sty1008}, \href
  {http://adsabs.harvard.edu/abs/2018MNRAS.478..651G} {478, 651}

\bibitem[\protect\citeauthoryear{{Gustafsson}, {Edvardsson}, {Eriksson},
  {J{\o}rgensen}, {Nordlund}  \& {Plez}}{{Gustafsson}
  et~al.}{2008}]{2008A&A...486..951G}
{Gustafsson} B.,  {Edvardsson} B.,  {Eriksson} K.,  {J{\o}rgensen} U.~G.,
  {Nordlund} {\AA}.,   {Plez} B.,  2008, \mn@doi [\aap]
  {10.1051/0004-6361:200809724}, \href
  {http://adsabs.harvard.edu/abs/2008A%26A...486..951G} {486, 951}

\bibitem[\protect\citeauthoryear{{Hamers}}{{Hamers}}{2017}]{Hamers2017}
{Hamers} A.~S.,  2017, \mn@doi [\mnras] {10.1093/mnras/stx035}, \href
  {http://adsabs.harvard.edu/abs/2017MNRAS.466.4107H} {466, 4107}

\bibitem[\protect\citeauthoryear{{Hayano} et~al.,}{{Hayano}
  et~al.}{2010}]{IRCSAO}
{Hayano} Y.,  et~al., 2010, in Adaptive Optics Systems II. p. 77360N,
  \mn@doi{10.1117/12.857567}

\bibitem[\protect\citeauthoryear{{Hidalgo} et~al.,}{{Hidalgo}
  et~al.}{2018}]{Hidalgo2018}
{Hidalgo} S.~L.,  et~al., 2018, \mn@doi [\apj] {10.3847/1538-4357/aab158},
  \href {http://adsabs.harvard.edu/abs/2018ApJ...856..125H} {856, 125}

\bibitem[\protect\citeauthoryear{{Hirano} et~al.,}{{Hirano}
  et~al.}{2016a}]{Hirano2016}
{Hirano} T.,  et~al., 2016a, \mn@doi [\apj] {10.3847/0004-637X/820/1/41}, \href
  {http://adsabs.harvard.edu/abs/2016ApJ...820...41H} {820, 41}

\bibitem[\protect\citeauthoryear{{Hirano} et~al.,}{{Hirano}
  et~al.}{2016b}]{2016ApJ...825...53H}
{Hirano} T.,  et~al., 2016b, \mn@doi [\apj] {10.3847/0004-637X/825/1/53}, \href
  {http://adsabs.harvard.edu/abs/2016ApJ...825...53H} {825, 53}

\bibitem[\protect\citeauthoryear{{H{\o}g} et~al.,}{{H{\o}g}
  et~al.}{2000}]{Tycho2}
{H{\o}g} E.,  et~al., 2000, \aap, \href
  {http://adsabs.harvard.edu/abs/2000A%26A...355L..27H} {355, L27}

\bibitem[\protect\citeauthoryear{{Howell} et~al.,}{{Howell}
  et~al.}{2014}]{Howell2014}
{Howell} S.~B.,  et~al., 2014, \mn@doi [\pasp] {10.1086/676406}, \href
  {http://adsabs.harvard.edu/abs/2014PASP..126..398H} {126, 398}

\bibitem[\protect\citeauthoryear{{Huang}, {Wu}  \& {Triaud}}{{Huang}
  et~al.}{2016}]{Huang2016}
{Huang} C.,  {Wu} Y.,   {Triaud} A.~H.~M.~J.,  2016, \mn@doi [\apj]
  {10.3847/0004-637X/825/2/98}, \href
  {http://adsabs.harvard.edu/abs/2016ApJ...825...98H} {825, 98}

\bibitem[\protect\citeauthoryear{{Johnson} et~al.,}{{Johnson}
  et~al.}{2018}]{2018MNRAS.481..596J}
{Johnson} M.~C.,  et~al., 2018, \mn@doi [\mnras] {10.1093/mnras/sty2238}, \href
  {https://ui.adsabs.harvard.edu/#abs/2018MNRAS.481..596J} {481, 596}

\bibitem[\protect\citeauthoryear{{Kobayashi} et~al.,}{{Kobayashi}
  et~al.}{2000}]{IRCS}
{Kobayashi} N.,  et~al., 2000, in {Iye} M.,  {Moorwood} A.~F.,  eds,  \procspie
  Vol. 4008, Optical and IR Telescope Instrumentation and Detectors. pp
  1056--1066, \mn@doi{10.1117/12.395423}

\bibitem[\protect\citeauthoryear{{Kreidberg}}{{Kreidberg}}{2015}]{Kreidberg2015}
{Kreidberg} L.,  2015, \mn@doi [\pasp] {10.1086/683602}, \href
  {http://adsabs.harvard.edu/abs/2015PASP..127.1161K} {127, 1161}

\bibitem[\protect\citeauthoryear{{Kuerster}, {Schmitt}, {Cutispoto}  \&
  {Dennerl}}{{Kuerster} et~al.}{1997}]{1997AA...320..831K}
{Kuerster} M.,  {Schmitt} J.~H.~M.~M.,  {Cutispoto} G.,   {Dennerl} K.,  1997,
  \aap, \href {http://adsabs.harvard.edu/abs/1997A%26A...320..831K} {320, 831}

\bibitem[\protect\citeauthoryear{{Kurucz}}{{Kurucz}}{1993}]{1993sssp.book.....K}
{Kurucz} R.~L.,  1993, {SYNTHE spectrum synthesis programs and line data}

\bibitem[\protect\citeauthoryear{{Laughlin}, {Marcy}, {Vogt}, {Fischer}  \&
  {Butler}}{{Laughlin} et~al.}{2005}]{Laughlin2005}
{Laughlin} G.,  {Marcy} G.~W.,  {Vogt} S.~S.,  {Fischer} D.~A.,   {Butler}
  R.~P.,  2005, \mn@doi [\apjl] {10.1086/444558}, \href
  {http://adsabs.harvard.edu/abs/2005ApJ...629L.121L} {629, L121}

\bibitem[\protect\citeauthoryear{{Lin}, {Bodenheimer}  \& {Richardson}}{{Lin}
  et~al.}{1996}]{Lin1996}
{Lin} D.~N.~C.,  {Bodenheimer} P.,   {Richardson} D.~C.,  1996, \mn@doi [\nat]
  {10.1038/380606a0}, \href {http://adsabs.harvard.edu/abs/1996Natur.380..606L}
  {380, 606}

\bibitem[\protect\citeauthoryear{{Livingston} et~al.,}{{Livingston}
  et~al.}{2018}]{2018AJ....156...78L}
{Livingston} J.~H.,  et~al., 2018, \mn@doi [\aj] {10.3847/1538-3881/aaccde},
  \href {https://ui.adsabs.harvard.edu/#abs/2018AJ....156...78L} {156, 78}

\bibitem[\protect\citeauthoryear{{Lundkvist} et~al.,}{{Lundkvist}
  et~al.}{2016}]{Lundkvist2016}
{Lundkvist} M.~S.,  et~al., 2016, \mn@doi [Nature Communications]
  {10.1038/ncomms11201}, \href
  {http://adsabs.harvard.edu/abs/2016NatCo...711201L} {7, 11201}

\bibitem[\protect\citeauthoryear{{Luri} et~al.,}{{Luri}
  et~al.}{2018}]{Luri2018}
{Luri} X.,  et~al., 2018, preprint, \href
  {http://adsabs.harvard.edu/abs/2018arXiv180409376L} {} (\mn@eprint {arXiv}
  {1804.09376})

\bibitem[\protect\citeauthoryear{{Mandel} \& {Agol}}{{Mandel} \&
  {Agol}}{2002}]{MandelAndAgol2002}
{Mandel} K.,  {Agol} E.,  2002, \mn@doi [\apjl] {10.1086/345520}, \href
  {http://adsabs.harvard.edu/abs/2002ApJ...580L.171M} {580, L171}

\bibitem[\protect\citeauthoryear{{Marshall}, {Robin}, {Reyl{\'e}}, {Schultheis}
   \& {Picaud}}{{Marshall} et~al.}{2006}]{2006A&A...453..635M}
{Marshall} D.~J.,  {Robin} A.~C.,  {Reyl{\'e}} C.,  {Schultheis} M.,   {Picaud}
  S.,  2006, \mn@doi [\aap] {10.1051/0004-6361:20053842}, \href
  {http://adsabs.harvard.edu/abs/2006A%26A...453..635M} {453, 635}

\bibitem[\protect\citeauthoryear{{Mayor} et~al.,}{{Mayor}
  et~al.}{2003}]{Mayor2003}
{Mayor} M.,  et~al., 2003, The Messenger, \href
  {http://adsabs.harvard.edu/abs/2003Msngr.114...20M} {114, 20}

\bibitem[\protect\citeauthoryear{{McLaughlin}}{{McLaughlin}}{1924}]{McLaughlin}
{McLaughlin} D.~B.,  1924, \mn@doi [\apj] {10.1086/142826}, \href
  {http://adsabs.harvard.edu/abs/1924ApJ....60...22M} {60}

\bibitem[\protect\citeauthoryear{{Mustill}, {Davies}  \& {Johansen}}{{Mustill}
  et~al.}{2015}]{2015ApJ...808...14M}
{Mustill} A.~J.,  {Davies} M.~B.,   {Johansen} A.,  2015, \mn@doi [\apj]
  {10.1088/0004-637X/808/1/14}, \href
  {http://adsabs.harvard.edu/abs/2015ApJ...808...14M} {808, 14}

\bibitem[\protect\citeauthoryear{{Pepe}, {Mayor}, {Galland}, {Naef}, {Queloz},
  {Santos}, {Udry}  \& {Burnet}}{{Pepe} et~al.}{2002}]{Pepe2002}
{Pepe} F.,  {Mayor} M.,  {Galland} F.,  {Naef} D.,  {Queloz} D.,  {Santos}
  N.~C.,  {Udry} S.,   {Burnet} M.,  2002, \mn@doi [\aap]
  {10.1051/0004-6361:20020433}, \href
  {https://ui.adsabs.harvard.edu/#abs/2002A&A...388..632P} {388, 632}

\bibitem[\protect\citeauthoryear{{Petrovich} \& {Tremaine}}{{Petrovich} \&
  {Tremaine}}{2016}]{Petrovich2016}
{Petrovich} C.,  {Tremaine} S.,  2016, \mn@doi [\apj]
  {10.3847/0004-637X/829/2/132}, \href
  {http://adsabs.harvard.edu/abs/2016ApJ...829..132P} {829, 132}

\bibitem[\protect\citeauthoryear{{Pu} \& {Lai}}{{Pu} \&
  {Lai}}{2018}]{2018MNRAS.478..197P}
{Pu} B.,  {Lai} D.,  2018, \mn@doi [\mnras] {10.1093/mnras/sty1098}, \href
  {http://adsabs.harvard.edu/abs/2018MNRAS.478..197P} {478, 197}

\bibitem[\protect\citeauthoryear{{Queloz} et~al.,}{{Queloz}
  et~al.}{2001}]{Queloz2001}
{Queloz} D.,  et~al., 2001, \mn@doi [\aap] {10.1051/0004-6361:20011308}, \href
  {http://adsabs.harvard.edu/abs/2001A%26A...379..279Q} {379, 279}

\bibitem[\protect\citeauthoryear{{Rafikov}}{{Rafikov}}{2006}]{Rafikov2006}
{Rafikov} R.~R.,  2006, \mn@doi [\apj] {10.1086/505695}, \href
  {http://adsabs.harvard.edu/abs/2006ApJ...648..666R} {648, 666}

\bibitem[\protect\citeauthoryear{{Rasio} \& {Ford}}{{Rasio} \&
  {Ford}}{1996}]{RasioFord1996}
{Rasio} F.~A.,  {Ford} E.~B.,  1996, \mn@doi [Science]
  {10.1126/science.274.5289.954}, \href
  {http://adsabs.harvard.edu/abs/1996Sci...274..954R} {274, 954}

\bibitem[\protect\citeauthoryear{{Ricker} et~al.,}{{Ricker}
  et~al.}{2014}]{TESS}
{Ricker} G.~R.,  et~al., 2014, in Space Telescopes and Instrumentation 2014:
  Optical, Infrared, and Millimeter Wave. p. 914320 (\mn@eprint {arXiv}
  {1406.0151}), \mn@doi{10.1117/12.2063489}

\bibitem[\protect\citeauthoryear{{Robin}, {Reyl{\'e}}, {Derri{\`e}re}  \&
  {Picaud}}{{Robin} et~al.}{2003}]{2003A&A...409..523R}
{Robin} A.~C.,  {Reyl{\'e}} C.,  {Derri{\`e}re} S.,   {Picaud} S.,  2003,
  \mn@doi [\aap] {10.1051/0004-6361:20031117}, \href
  {http://adsabs.harvard.edu/abs/2003A%26A...409..523R} {409, 523}

\bibitem[\protect\citeauthoryear{{Rossiter}}{{Rossiter}}{1924}]{Rossiter}
{Rossiter} R.~A.,  1924, \mn@doi [\apj] {10.1086/142825}, \href
  {http://adsabs.harvard.edu/abs/1924ApJ....60...15R} {60}

\bibitem[\protect\citeauthoryear{{Schwarz}, {Funk}, {Zechner}  \&
  {Bazs{\'o}}}{{Schwarz} et~al.}{2016}]{Schwarz2016}
{Schwarz} R.,  {Funk} B.,  {Zechner} R.,   {Bazs{\'o}} {\'A}.,  2016, \mn@doi
  [\mnras] {10.1093/mnras/stw1218}, \href
  {http://adsabs.harvard.edu/abs/2016MNRAS.460.3598S} {460, 3598}

\bibitem[\protect\citeauthoryear{{Silva Aguirre} et~al.,}{{Silva Aguirre}
  et~al.}{2015}]{Victor2015}
{Silva Aguirre} V.,  et~al., 2015, \mn@doi [\mnras] {10.1093/mnras/stv1388},
  \href {http://adsabs.harvard.edu/abs/2015MNRAS.452.2127S} {452, 2127}

\bibitem[\protect\citeauthoryear{{Telting} et~al.,}{{Telting}
  et~al.}{2014}]{FIES2014}
{Telting} J.~H.,  et~al., 2014, \mn@doi [Astronomische Nachrichten]
  {10.1002/asna.201312007}, \href
  {http://cdsads.u-strasbg.fr/abs/2014AN....335...41T} {335, 41}

\bibitem[\protect\citeauthoryear{{Thorngren}, {Fortney}, {Murray-Clay}  \&
  {Lopez}}{{Thorngren} et~al.}{2016}]{Thorngren2016}
{Thorngren} D.~P.,  {Fortney} J.~J.,  {Murray-Clay} R.~A.,   {Lopez} E.~D.,
  2016, \mn@doi [\apj] {10.3847/0004-637X/831/1/64}, \href
  {http://adsabs.harvard.edu/abs/2016ApJ...831...64T} {831, 64}

\bibitem[\protect\citeauthoryear{{Torres}}{{Torres}}{2010}]{2010AJ....140.1158T}
{Torres} G.,  2010, \mn@doi [\aj] {10.1088/0004-6256/140/5/1158}, \href
  {http://adsabs.harvard.edu/abs/2010AJ....140.1158T} {140, 1158}

\bibitem[\protect\citeauthoryear{{Van Eylen} \& {Albrecht}}{{Van Eylen} \&
  {Albrecht}}{2015}]{VanEylen2015}
{Van Eylen} V.,  {Albrecht} S.,  2015, \mn@doi [\apj]
  {10.1088/0004-637X/808/2/126}, \href
  {http://adsabs.harvard.edu/abs/2015ApJ...808..126V} {808, 126}

\bibitem[\protect\citeauthoryear{{Van Eylen} et~al.,}{{Van Eylen}
  et~al.}{2018a}]{VanEylen2019}
{Van Eylen} V.,  et~al., 2018a, preprint, \href
  {http://adsabs.harvard.edu/abs/2018arXiv180700549V} {} (\mn@eprint {arXiv}
  {1807.00549})

\bibitem[\protect\citeauthoryear{{Van Eylen} et~al.,}{{Van Eylen}
  et~al.}{2018b}]{2018MNRAS.478.4866V}
{Van Eylen} V.,  et~al., 2018b, \mn@doi [\mnras] {10.1093/mnras/sty1390}, \href
  {https://ui.adsabs.harvard.edu/#abs/2018MNRAS.478.4866V} {478, 4866}

\bibitem[\protect\citeauthoryear{{Van Eylen}, {Agentoft}, {Lundkvist},
  {Kjeldsen}, {Owen}, {Fulton}, {Petigura}  \& {Snellen}}{{Van Eylen}
  et~al.}{2018c}]{VanEylen2018}
{Van Eylen} V.,  {Agentoft} C.,  {Lundkvist} M.~S.,  {Kjeldsen} H.,  {Owen}
  J.~E.,  {Fulton} B.~J.,  {Petigura} E.,   {Snellen} I.,  2018c, \mn@doi
  [\mnras] {10.1093/mnras/sty1783}, \href
  {http://adsabs.harvard.edu/abs/2018MNRAS.479.4786V} {479, 4786}

\bibitem[\protect\citeauthoryear{{Vanderburg} \& {Johnson}}{{Vanderburg} \&
  {Johnson}}{2014}]{Vanderburg2014}
{Vanderburg} A.,  {Johnson} J.~A.,  2014, \mn@doi [\pasp] {10.1086/678764},
  \href {http://adsabs.harvard.edu/abs/2014PASP..126..948V} {126, 948}

\bibitem[\protect\citeauthoryear{{Weiss} \& {Marcy}}{{Weiss} \&
  {Marcy}}{2014}]{Weiss2014}
{Weiss} L.~M.,  {Marcy} G.~W.,  2014, \mn@doi [\apjl]
  {10.1088/2041-8205/783/1/L6}, \href
  {http://adsabs.harvard.edu/abs/2014ApJ...783L...6W} {783, L6}

\bibitem[\protect\citeauthoryear{{Winn}, {Fabrycky}, {Albrecht}  \&
  {Johnson}}{{Winn} et~al.}{2010}]{Winn2010}
{Winn} J.~N.,  {Fabrycky} D.,  {Albrecht} S.,   {Johnson} J.~A.,  2010, \mn@doi
  [\apjl] {10.1088/2041-8205/718/2/L145}, \href
  {http://adsabs.harvard.edu/abs/2010ApJ...718L.145W} {718, L145}

\bibitem[\protect\citeauthoryear{{Zanazzi} \& {Lai}}{{Zanazzi} \&
  {Lai}}{2018}]{ZanazziAndLai2018}
{Zanazzi} J.~J.,  {Lai} D.,  2018, \mn@doi [\mnras] {10.1093/mnras/sty951},
  \href {http://adsabs.harvard.edu/abs/2018MNRAS.477.5207Z} {477, 5207}

\bibitem[\protect\citeauthoryear{{Zechmeister} \& {K{\"u}rster}}{{Zechmeister}
  \& {K{\"u}rster}}{2009}]{LS}
{Zechmeister} M.,  {K{\"u}rster} M.,  2009, \mn@doi [\aap]
  {10.1051/0004-6361:200811296}, \href
  {http://adsabs.harvard.edu/abs/2009A%26A...496..577Z} {496, 577}

\makeatother
\end{thebibliography}


\appendix

\textcolor{black}{
\section{Input for the Besan\c{c}on model}
\label{appendix_besancon}
The {\tt Besan\c{c}on} Galactic population model \citep{2003A&A...409..523R} is initialized at a 1~deg$^2$ area centered on the galactic coordinates of star A ($l=348.0523$~deg, $b=+27.5996$~deg). We do the calculations without kinematics and use the dust map of \citet{2006A&A...453..635M} assuming no dispersion on the extinction. With these settings we calculate the number of background sources in a 10~kpc radius brighter than $H=15$, which safely encompasses errors on the $H$-magnitude of star B. This is used to estimate the chance alignment probability in Sec. \ref{subsec:companions}.
}

\section{Extra material}

\begin{figure*}
\centering
\includegraphics[width=0.99\textwidth]{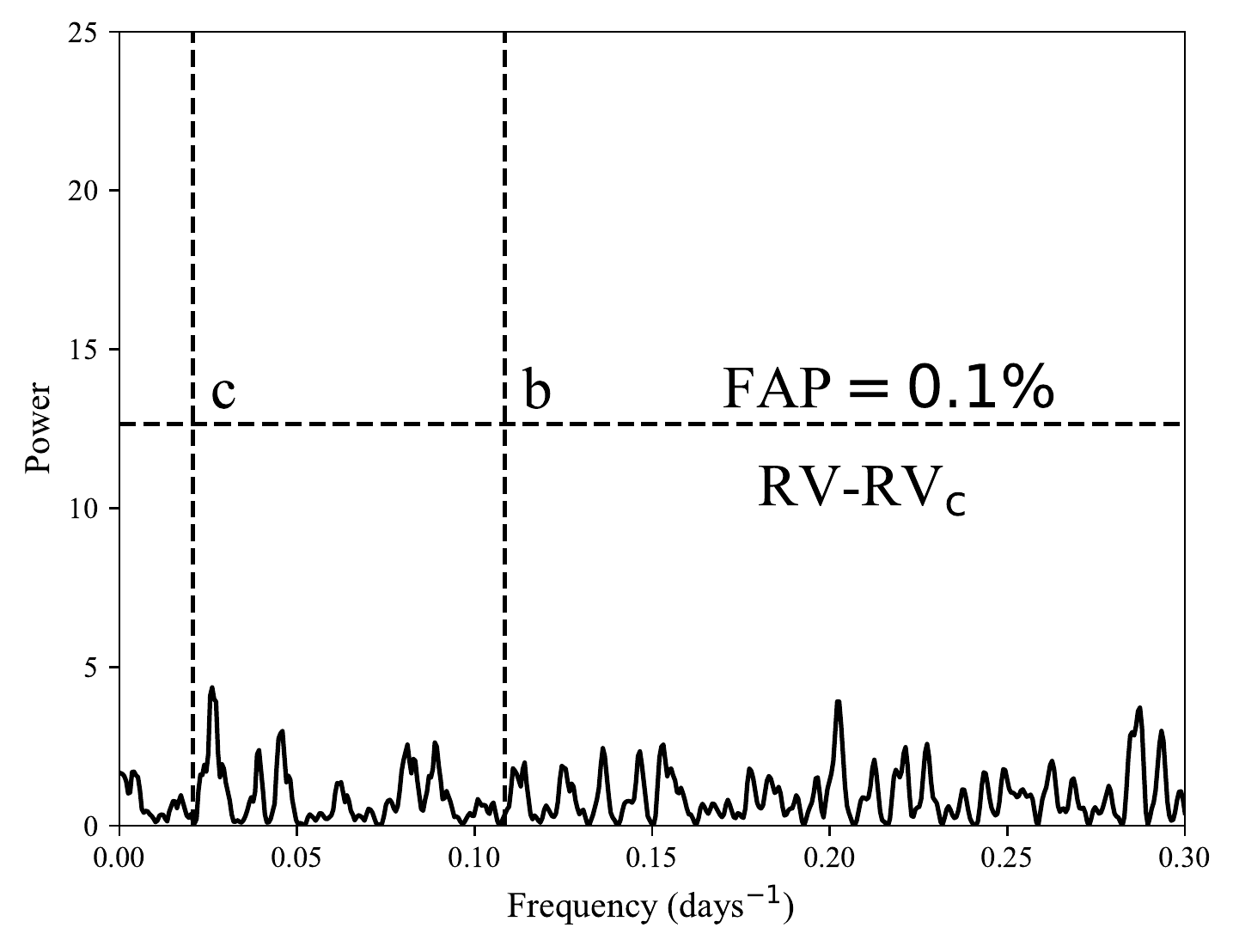}
\hspace*{-1cm}
\mbox{\llap{\raisebox{8.5cm}{\includegraphics[width=0.34\textwidth]{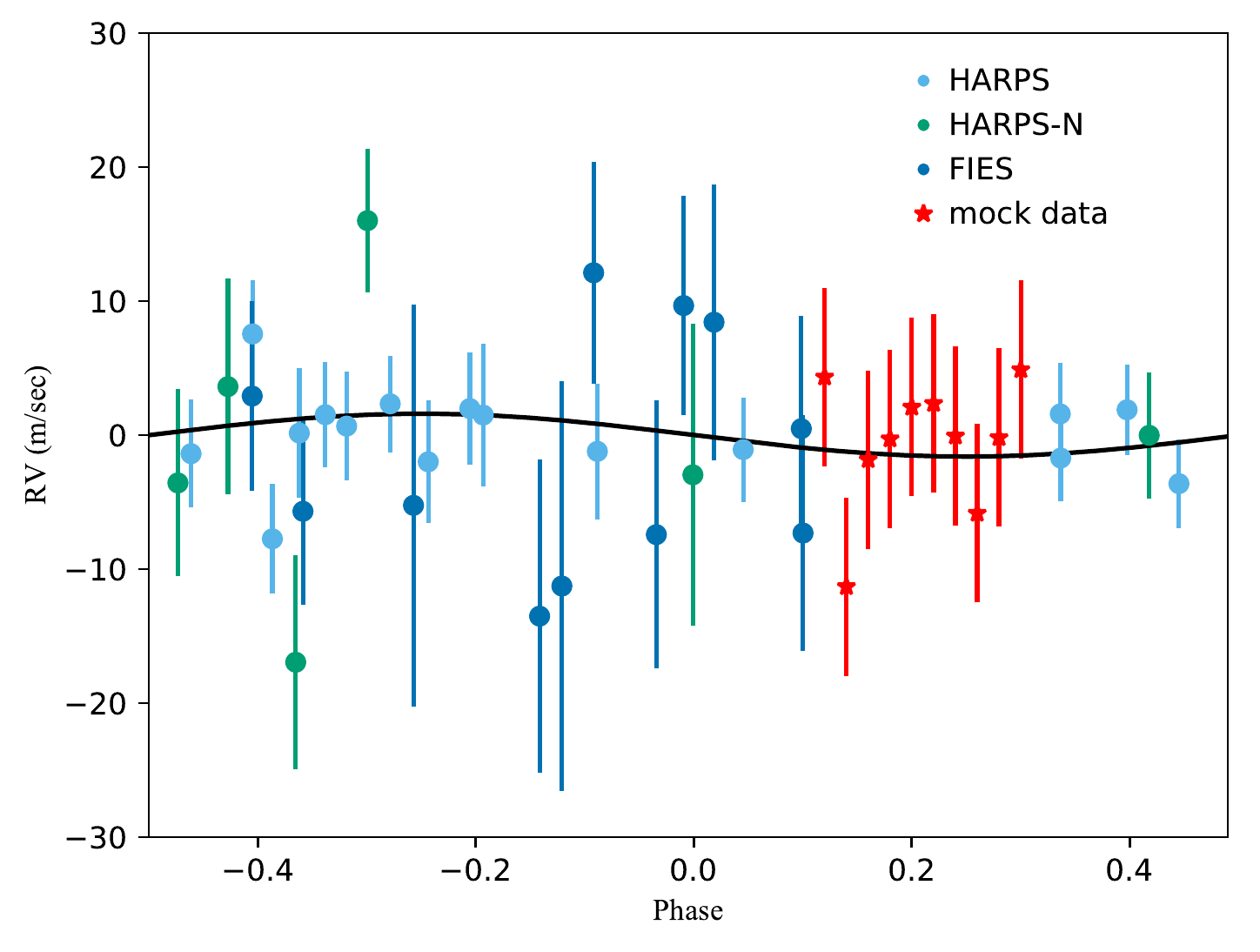}}}}
\hspace{1cm}
\caption{\textcolor{black}{Same as the right part of figure \ref{fig:LS}, but where the frequency analysis includes simulated data in planet b's spotty phase region of 0.1-0.3. The mock data were created by adding Gaussian noise equal to the mean noise value of the real data, $6.7$~m~s$^{-1}$, to a Keplerian model. The simulated data in this specific LS periodogram is done with $K=1.6$~m~s$^{-1}$, and can be seen as the inset in the right corner (with this specific Keplerian model shown as a black line). The analysis was repeatedly performed for values of K up to the upper limit of 6.6~m~s$^{-1}$. In neither of these, the planet was detected above the FAP treshold of $0.1\%$.}}
\label{fig:LS_mockdata}
\end{figure*}

\begin{table*}
\begin{center}
\centering
\caption{Radial velocities and related values for K2-290 using the HARPS, HARPS-N and FIES spectrographs. We list the barycentric time of mid-exposure, the RVs, the instrumental RV uncertainties ($\sigma_{RV}$), the bisector span (BIS) and the FWHM of the CCFs, the exposure times (t$_{\text{exp}}$), the signal-to-noise ratios (S/N), and the instrument used for a specific observation.
\textbf{Notes.}
$^{*}$S/N is per pixel and is calculated at 5500 \AA.
}
\label{table:RVs}
\begin{tabular}{cccccccc}
\hline\hline
\noalign{\smallskip}
Time (BJD$_{\text{TDB}}$) & RV-19700 (m s$^{-1}$) & $\sigma_{RV}$ (m s$^{-1}$) & BIS (m s$^{-1}$) & FWHM (km s$^{-1}$) & t$_{\text{exp}}$ (s) & S/N* & Instr.\\
\hline
\noalign{\smallskip}
2458172.894740 & 21.7 & 3.2 & 55.1 & 10.9581 & 1800 & 72.3 & HARPS\\
2458175.890871 & 23.9 & 3.9 & 35.0 & 10.9632 & 1500 & 60.8 & HARPS\\
2458191.878876 & 73.0 & 3.4 & 35.3 & 10.9514 & 2100 & 68.6 & HARPS\\
2458193.866635 & 71.5 & 4.1 & 44.4 & 10.9822 & 1500 & 53.4 & HARPS\\
2458194.862556 & 85.6 & 3.6 & 53.7 & 10.9684 & 1800 & 65.3 & HARPS\\
2458197.846021 & 93.1 & 3.9 & 53.7 & 10.9733 & 1800 & 59.2 & HARPS\\
2458220.814348 & 22.7 & 4.0 & 45.4 & 10.9481 & 2400 & 56.6 & HARPS\\
2458221.728938 & 23.0 & 4.8 & 60.5 & 10.9637 & 1800 & 46.7 & HARPS\\
2458222.818965 & 20.2 & 4.6 & 34.4 & 10.9358 & 1800 & 50.9 & HARPS\\
2458249.764215 & 103.6 & 4.1 & 44.9 & 10.9486 & 2300 & 57.5 & HARPS\\
2458250.805571 & 106.0 & 4.2 & 47.6 & 10.9872 & 2100 & 55.9 & HARPS\\
2458324.611686 & 29.1 & 5.3 & 9.7 & 10.9769 & 2400 & 45.1 & HARPS\\
2458325.576369 & 28.6 & 5.1 & 39.6 & 10.9385 & 2400 & 47.4 & HARPS\\
2458329.488931 & 43.2 & 3.8 & 34.4 & 10.9725 & 2100 & 58.5 & HARPS\\
2458330.491675 & 41.6 & 3.3 & 42.0 & 10.9761 & 2100 & 68.6 & HARPS\\
2458359.508754 & 58.6 & 4.0 & 38.6 & 10.9721 & 2400 & 56.7 & HARPS\\
2458169.785818 & 28.2 & 11.2 & 68.0 & 11.0293 & 1200 & 25.7 & HARPS-N\\
2458202.702616 & 107.7 & 8.0 & 83.0 & 10.9708 & 3600 & 34.0 & HARPS-N\\
2458219.700532 & 26.3 & 4.7 & 60.4 & 10.9347 & 2100 & 52.8 & HARPS-N\\
2458220.703092 & 20.6 & 7.0 & 61.0 & 10.9395 & 2100 & 38.0 & HARPS-N\\
2458221.698890 & 5.8 & 8.0 & 95.4 & 10.9209 & 2100 & 33.6 & HARPS-N\\
2458314.421992 & 49.0 & 5.4 & 42.4 & 10.9777 & 1800 & 47.2 & HARPS-N\\
2458251.583280 & -36.5 & 15.3 & 46.8 & 16.8868 & 3600 & 32.2 & FIES\\
2458253.621643 & -34.5 & 8.8 & 34.6 & 16.7709 & 3000 & 54.5 & FIES\\
2458258.606573 & -55.3 & 7.0 & 44.4 & 16.9018 & 3600 & 62.7 & FIES\\
2458259.540916 & -61.2 & 15.0 & 11.5 & 16.8045 & 3600 & 31.2 & FIES\\
2458260.606755 & -77.0 & 11.7 & 34.1 & 16.8718 & 3600 & 48.5 & FIES\\
2458261.593343 & -77.9 & 10.0 & 21.5 & 16.8944 & 3600 & 49.2 & FIES\\
2458279.486551 & -81.2 & 8.3 & 96.9 & 16.8609 & 3600 & 59.0 & FIES\\
2458280.504146 & -81.6 & 10.3 & 28.3 & 16.8988 & 3600 & 49.5 & FIES\\
2458289.458960 & -45.3 & 8.2 & 48.9 & 16.8567 & 3600 & 61.1 & FIES\\
2458290.453468 & -50.4 & 8.4 & 32.1 & 16.8290 & 3600 & 59.9 & FIES\\
2458313.447528 & -89.0 & 7.1 & 34.7 & 16.8713 & 3600 & 58.3 & FIES
\end{tabular}
\end{center}
\end{table*}

\begin{table*}
\begin{center}
\centering
\caption{\textcolor{black}{Same as Table \ref{table:system_parameters}, but with the eccentric solution of planet c's orbit. {\bf{Notes:}} *Because of $\omega=70.0\pm 9.0$~deg being close to $90$~deg, we regard this solution as highly suspicious (se Sec. \ref{planet_parameters}).}}
\label{table:system_parameters_ecc}
\begin{tabular}{lc}
\hline\hline
\noalign{\smallskip}
Parameters from RV and transit MCMC analysis                         & Planet c (eccentric)* \\
\noalign{\smallskip}
\hline
Quadratic limb darkening parameter $c_1$                             & $0.329\pm 0.037$                                       \\
Quadratic limb darkening parameter $c_2$                             & $0.219\pm 0.067$                                       \\
Noise term K2 $\sigma_{\text{K2}}$                             & $0.0000209^{+0.0000044}_{-0.0000052}$                                       \\
Jitter term FIES $\sigma_{\text{jit,FIES}}$ (m s$^{-1}$)         & $4.1_{-2.8}^{+4.4}$                                       \\
Jitter term HARPS $\sigma_{\text{jit,HARPS}}$ (m s$^{-1}$)       & $1.5_{-1.0}^{+1.6}$                                       \\
Jitter term HARPS-N $\sigma_{\text{jit,HARPS-N}}$ (m s$^{-1}$)   & $11.1_{-4.8}^{+7.2}$                                       \\
Systemic velocity FIES $\gamma_{\text{FIES}}$ (km s$^{-1}$)       & $19.6316_{-0.0033}^{+0.0032}$                                       \\
Systemic velocity HARPS $\gamma_{\text{HARPS}}$ (km s$^{-1}$)     & $19.7612\pm 0.0013$                                       \\
Systemic velocity HARPS-N $\gamma_{\text{HARPS-N}}$ (km s$^{-1}$) & $19.7611_{-0.0059}^{+0.0057}$                                       \\
Orbital period $P$ (days)                                        & $48.36692^{+0.00040}_{-0.00042}$                            \\
Time of midttransit $T_0$ (BJD)                             & $2458019.17336\pm 0.00029$                            \\
Scaled planetary radius $R_{\text{p}}/R_{\star}$                     & $0.06758\pm 0.00057$                            \\
Scaled orbital distance $a/R_{\star}$                                & $40.1\pm 1.5$                            \\
Orbital inclination $i$ (deg)                                    & $89.41^{+0.17}_{-0.14}$                            \\
RV semi-amplitude $K_{\star}$ (m s$^{-1}$)                       & $41.1 \pm 1.7$                            \\
$\sqrt{e} \sin(\omega)$                                          & $0.354^{+0.043}_{-0.050}$                            \\
$\sqrt{e} \cos(\omega)$                                          & $0.130^{+0.052}_{-0.059}$                            \\
\hline
\noalign{\smallskip}
Derived parameters                                                   & \multicolumn{1}{l}{}         \\
\noalign{\smallskip}
\hline
Orbital eccentricity $e$                                             & $0.144^{+0.033}_{-0.032}$                            \\
Argument of periastron $\omega$ (deg)                            & $70.0\pm 9.0$                          \\
Impact parameter $b$                                             & \textcolor{black}{$0.358\pm 0.018$}                            \\
Total transit duration $T_{14}$ (hr)                           & $8.09 \pm 0.47$                            \\
Full transit duration \textcolor{black}{$T_{23}$} (hr)                   & $6.92\pm 0.46$                            \\
Planetary mass $M_{\text{p}}$ ($M_{\text{J}}$)                   & $0.819\pm 0.048$                           \\
Planetary radius $R_{\text{p}}$ ($R_{\text{J}}$)                 & $0.993\pm 0.050$                            \\
Planetary mean density $\rho_{\text{p}}$ (g cm$^{-3}$)                & $1.11\pm 0.18$                            \\
semi-major axis a (AU)                                           & $0.281\pm 0.017$                            \\
Equlibrium temperature $T_{\text{eq}}$ (K)                       & $704\pm 19$                            \\

\end{tabular}
\end{center}
\end{table*}

\label{lastpage}
\end{document}